\documentclass[iop,apj,revtex4]{emulateapj}

\begin{document}

\title{Probing cosmological isotropy with type Ia Supernovae}
\author{C.A.P. Bengaly Jr., A. Bernui, J.S. Alcaniz
}
\affil{Observat\'orio Nacional, 20921-400, Rio de Janeiro - RJ, Brasil}

\begin{abstract}
We investigate the validity of the Cosmological Principle by mapping the cosmological parameters  $H_0$ and $q_0$ through the celestial sphere. In our analysis, performed in a low-redshift regime to follow a model-independent approach, we use two compilations of type Ia Supernovae (SNe Ia), namely the Union2.1 and the JLA datasets. 
Firstly, we show that the angular distributions for both SNe Ia datasets are statistically anisotropic at high confidence level ($p$-value $<$ 0.0001), in particular the JLA sample. Then we find that the cosmic expansion and acceleration are mainly of dipolar type, with maximal anisotropic expansion [acceleration] pointing towards $(l,b) \simeq (326^{\circ},12^{\circ})$ [$(l,b) \simeq (174^{\circ},27^{\circ})$], and $(l,b) \simeq (58^{\circ},-60^{\circ})$ [$(l,b) \simeq (225^{\circ},51^{\circ})$] for the Union2.1 and JLA data, respectively. 
Secondly, we use a geometrical method to test the hypothesis that the non-uniformly distributed SNe Ia events could introduce anisotropic imprints on the cosmological expansion and acceleration. 
For the JLA compilation, we found significant correlations between the celestial distribution of data points and the directional studies of $H_0$ and $q_0$, suggesting that these results can be attributed to the intrinsic anisotropy of the sample. In the case of the Union2.1 data, nonetheless, these correlations are less pronounced, and we verify that the dipole asymmetry found in the $H_0$ analyses coincides with the well-known bulk-flow motion of our local group. 
From these analyses, we conclude that the directional asymmetry on the cosmological parameters maps are mainly either of local origin or due to celestial incompleteness of current SNe Ia samples.
\end{abstract} 

\date{\today}

\section{Introduction}  \label{intro}

Spatial homogeneity and isotropy are large-scale properties incorporated into the standard description of the Universe via the Cosmological Principle (CP) (see, e.g., Goodman, 1995; Maartens, 2011; Clarkson, 2012 for a discussion). According to current cosmological data (Eisenstein {\it et al.}, 2005; Blake {\it et al.} 2011; Suzuki {\it et al.}, 2012; Bennett {\it et al.}, 2013; Ade {\it et al.}, 2013;  Betoule {\it et al.}, 2014), the best picture nowadays is provided by the  so-called $\Lambda$CDM model, a homogeneous and isotropic scenario which accounts well for the current accelerating phase of the Universe, most of the physical features in the Cosmic Microwave Background (CMB) spectrum and the observed large-scale structure. 

From the theoretical viewpoint, however, it is well known that there is still no satisfactory description of the mechanism behind cosmic acceleration which is a crucial aspect to the cosmological modelling. This in turn motivates the need to probe fundamental hypotheses in Cosmology which includes the validity of the CP. Moreover, recent analyses have reported cosmological anomalies that may indicate deviations from an  isotropic Universe, such as large scale bulk-flow (Colin {\it et al.}, 2011, Turnbull {\it et al.}, 2012, Feindt {\it et al.}, 2013), alignment of low multipoles in the CMB power spectrum, CMB hemispherical asymmetry (Eriksen {\it et al.}, 2004; Bernui {\it et al.}, 2006), large scale alignment in the optical polarization QSO data (Hutsem\'ekers {\it et al.}, 2005; Hutsem\'ekers {\it et al.}, 2014) and spatial dependence of the value of the fine structure constant (Mariano \& Perivolaropoulos, 2012). If an isotropic expansion is observationally confirmed, it would reinforce the search for a 
cosmological description in the Friedmann-Lema\^itre-Robertson-Walker class. Otherwise, models with a more general spatial geometry should be seriously explored to describe the structure and evolution of the Universe.  

The idea of testing the CP with the Hubble diagram is recent. Previous efforts were done with several catalogs (Kolatt \& Lahav, 2001; Schwarz \& Weinhorst, 2007; Antoniou \& Perivolaropoulos, 2010; Gumpta \& Saini, 2010; Cai \& Tuo, 2012; Kalus {\it et al.}, 2013; Zhao {\it et al.}, 2013; Chang {\it et al.}, 2014a, Gupta \& Singh, 2014; Chang \& Lin, 2015; Jim\'enez {\it et al.}, 2015), where the authors attempted to constrain the cosmological isotropy with the opposite hemispheres technique, or assuming 
anisotropic dark energy models (Blomqvist {\it et al.}, 2010; Mariano \& Perivolaropoulos, 2012; Cai {\it et al.}, 2013; Yang {\it et al.}, 2014;  Chang {\it et al.}, 2014b; Wang {\it et al.}, 2014). These efforts reported the validity of the CP within 2$\sigma$ confidence level, with the dipolar directions of maximal anisotropy pointing in the direction of the cosmic anomalies we have mentioned earlier.  

The novelty of this work is a directional analysis of 
the cosmological parameters $H_0$ and $q_0$ in different sky patches by means of hemispherical comparison, including the investigation whether possible anisotropic signals in such analyses can be correlated with the non-uniform distribution of SNe events in the sky (according to the current catalogs). If a significant correlation is found, this would indicate that such signal could be attributed to observational biases rather than being of cosmological nature. We use the two latest compilations of SNe in order to perform these analyses, namely the Union2.1 (Suzuki {\it et al.}, 2012) and the JLA datasets (Betoule {\it et al.}, 2014) in a reasonably low-redshift regime, i.e., $z \leq 0.20$. This choice has been made as there are approximately one 
third of the objects in both samples located within this range of redshift and also because of the advantage of carrying out model-independent analyses as pointed out by Seikel \& Schwarz (2009) and Kalus {\it et al.} (2013).
We use galactic coordinates throughout all our analyses. 

 
\vskip 1.5cm
 
\section{Some methods to study statistical isotropy in SNe data} \label{methods}

\subsection{Sigma-map} \label{sigma-map}

In this section we describe our indicator and the procedure to calculate their associated maps, which lead to quantify deviations from statistical isotropy in a given set of cosmic events with known positions on the celestial sphere (Bernui {\it et al.}, 2008; see also Bernui {\it et al.}, 2007). Our primary purpose is to illustrate the procedure for defining the discrete function $\sigma$ on 
the celestial sphere in order to generate their associated maps, called $\sigma-$maps, which 
compared with statistically isotropic simulated maps give us a measure of deviations from 
isotropy in the data set. 

Let $\Omega_j^{\gamma_0} \equiv \Omega(\theta_j,\phi_j;\gamma_0) \in {\cal S}^2$ be a 
spherical cap region on the celestial sphere, of $\gamma_0$ degrees of aperture, with centre at 
the $j$-th pixel, $j=1, \ldots, N_{\mbox{\tiny caps}}$, where $(\theta_j,\phi_j)$ are the 
angular coordinates of the center of the $j$-th pixel. Both, the number of spherical caps 
$N_{\mbox{\tiny caps}}$ and the coordinates of their center $(\theta_j,\phi_j)$ are 
defined using the HEALPix pixelization scheme (G\'orski {\it et al.}, 2005). 
The spherical caps are such that their union completely covers the celestial sphere ${\cal S}^2$.

Let ${\cal{C}}^{\,j}$ be the catalog of cosmic objects located in the $j$-th spherical cap 
$\Omega_j^{\gamma_0}$. 
The 2PACF of these objects (Padmanabhan, 1993), denoted as $\Delta_j(\gamma_i;\gamma_0)$, 
is the difference between the normalised frequency distribution and that expected from the 
number of pairs-of-objects with angular distances in the 
interval $(\gamma_i - 0.5\delta,\gamma_i + 0.5\delta],\, i=1,\ldots,N_{\mbox{\tiny bins}}$, 
where $\gamma_i \equiv (i-0.5)\delta$ and $\delta \equiv 2\gamma_0 / N_{\mbox{\tiny bins}}$ 
is the bin width. 
The expected distribution is the average of normalised frequency distributions obtained from a 
large number of simulated realisations of isotropically distributed objects in ${\cal S}^2$, 
containing a similar number of objects as in the dataset in analysis. 
A positive (negative) value of $\Delta_j$ indicates that objects with these angular separations 
are correlated (anti-correlated), while zero indicates no correlation. 

Let us define now the scalar function 
$\sigma: \Omega_j^{\gamma_0} \mapsto {\Re}^{+}$, for $j=1, \ldots, N_{\mbox{\tiny caps}}$, 
which assigns to the $j$-cap, centered at $(\theta_j,\phi_j)$, a real positive number 
$\sigma_j \equiv \sigma(\theta_j,\phi_j) \in \Re^+$. 
We define a measure $\sigma$ of the angular correlations in the $j$-cap as 
\begin{equation} \label{sigma}
\sigma^2_j  \equiv \frac{1}{N_{\mbox{\tiny bins}}}
\sum_{i=1}^{N_{\mbox{\tiny bins}}} \Delta^2_j (\gamma_i;\gamma_0).
\end{equation}

\noindent
To obtain a quantitative measure of the angular correlation signatures of the SNe sky map, 
we choose $\gamma_0 = 90^{\circ}$ and cover the celestial sphere with 
$N_{\mbox{\tiny caps}}=N_{\mbox{\tiny hems}}=768$ hemispheres, then calculate the set of 
values $\{ \sigma_j, \, j=1,...,N_{\mbox{\tiny caps}} \}$ using eq.~(\ref{sigma}). 
Patching together the set $\{ \sigma_j \}$ in the celestial sphere  according to a coloured 
scale (where, for instance, $\sigma^{\mbox{\footnotesize minimal}} \rightarrow blue$, 
$\sigma^{\mbox{\footnotesize maximal}} \rightarrow red$) we obtain a sigma-map. 
Finally, we quantify the angular correlation signatures of a given sigma-map by calculating its 
angular power spectrum. 
Similar power spectra are constructed, for comparison, with isotropically distributed samples 
of cosmic objects.


\begin{figure*}
\includegraphics[width = 6.0cm, height = 8.75cm, angle = +90]{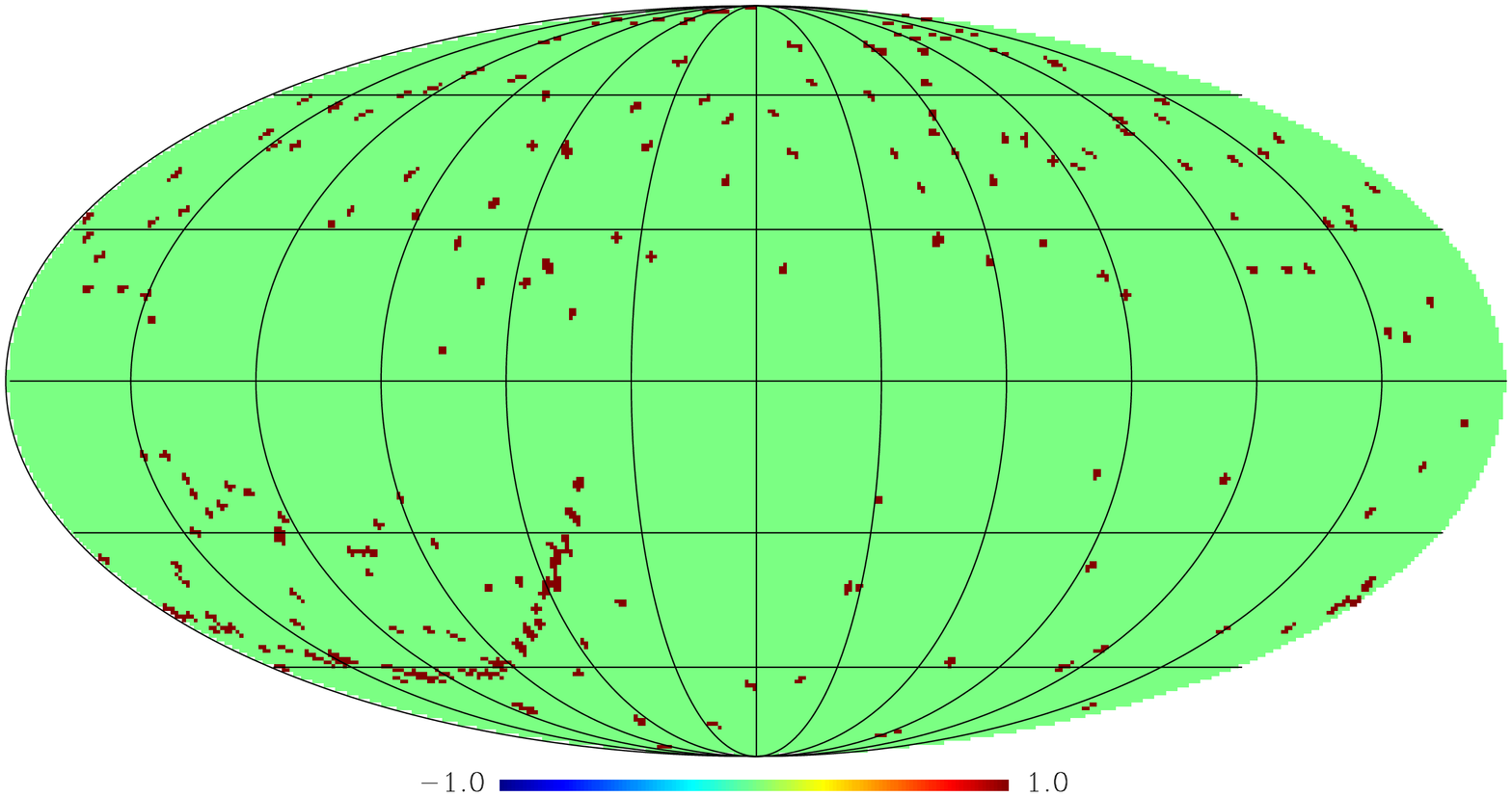}
\hspace{0.3cm}
\includegraphics[width = 6.0cm, height = 8.75cm, angle = +90]{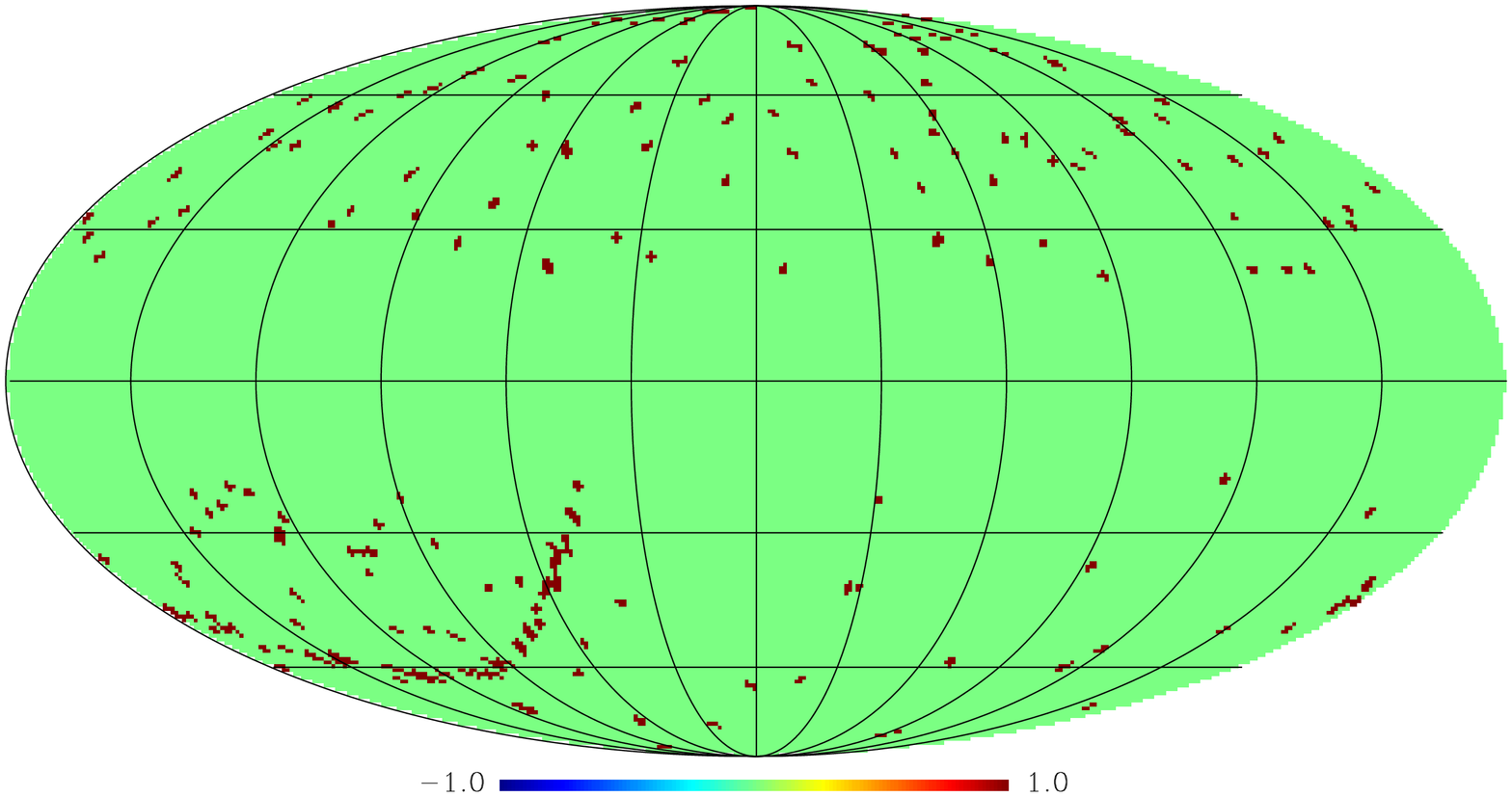}
\caption{ {\it Left panel:} The Union2.1 SNe distribution in the celestial sphere for $z \le 0.20$ (230 data points). 
{\it Right panel:} The same catalog with the mask20 applied. 19 SNe have been 
excluded in this cut.} 
\label{fig1}
\end{figure*}



\begin{figure*}
\includegraphics[width = 6.0cm, height = 8.75cm, angle = +90]{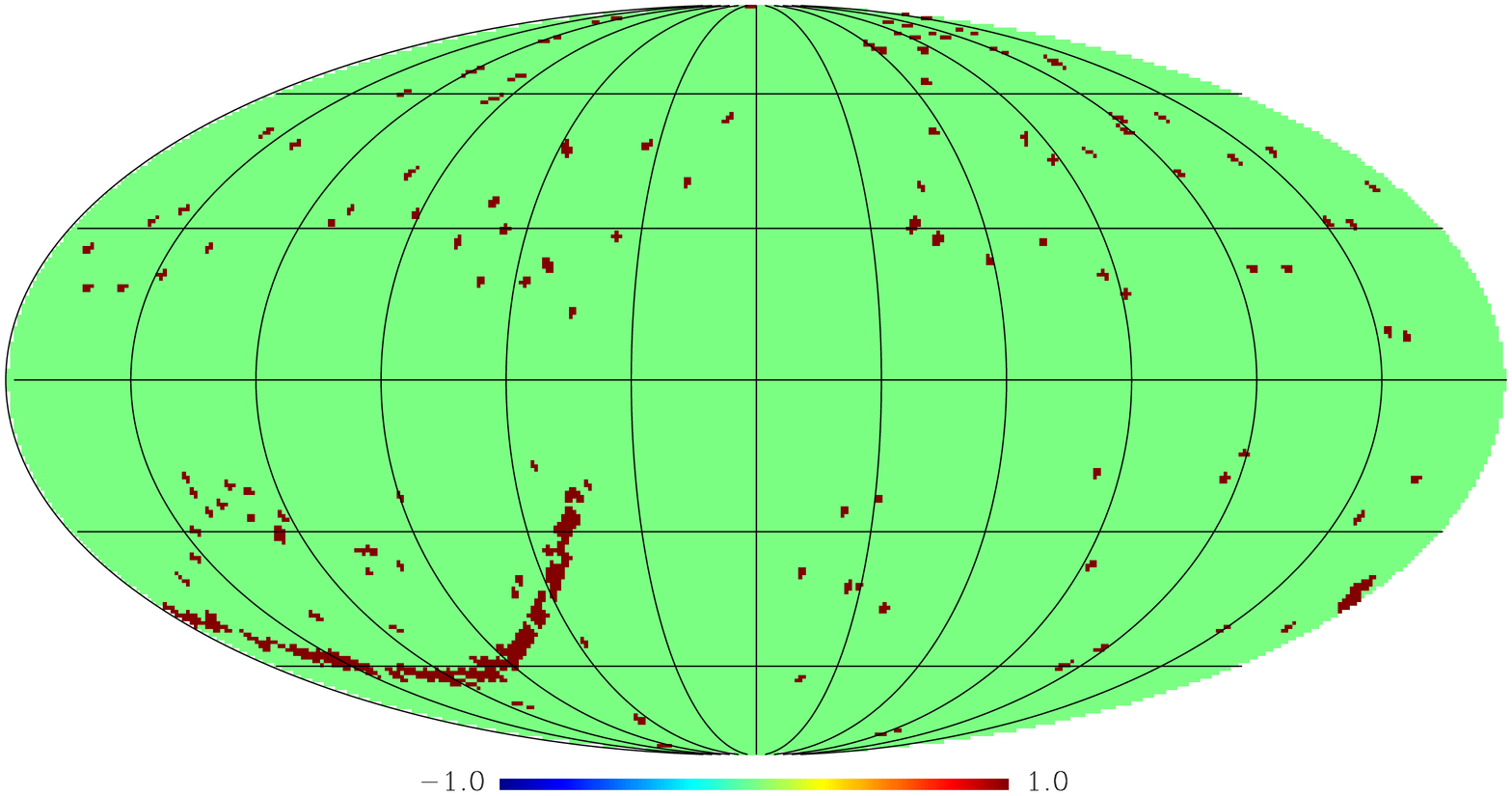}
\hspace{0.3cm}
\includegraphics[width = 6.0cm, height = 8.75cm, angle = +90]{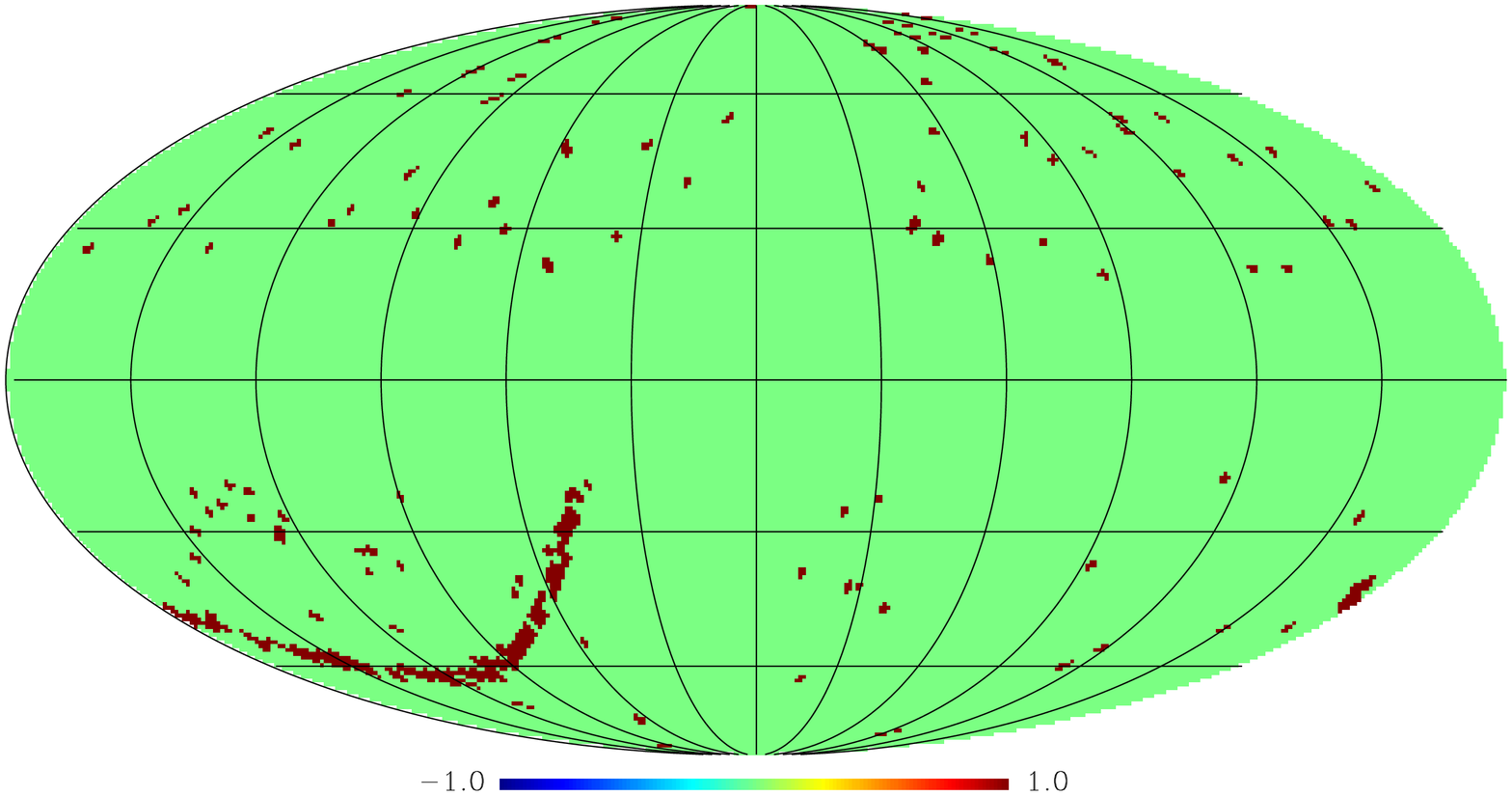}
\caption{ {\it Left panel:} The Union2.1 SNe distribution in the celestial sphere for $z \le 0.20$ (317 data points). 
{\it Right panel:} The same catalog with the mask20 applied. 15 SNe have been excluded in this cut.} 
\label{fig2}
\end{figure*}


\subsection{Spherical harmonics expansion} \label{c_eles}

Since the sigma-map assigns a real number value to each pixel in the celestial sphere, that is 
$\sigma = \sigma(\theta,\phi)$, one can expand it in spherical harmonics: 
$\sigma(\theta,\phi) = \sum_{\ell,\, m} A_{\ell\, m} Y_{\ell\, m}(\theta,\phi)$ where the set of 
values $\{ C_{\ell} \}$, 
defined by $C_{\ell} \equiv (1 / (2\ell+1)) \sum_{m={\mbox{\small -}}\ell}^{\ell} \, |A_{\ell\, m}|^2$, 
is the angular power spectrum of the sigma-map. Because we are interested in the large-scale 
angular correlations, we shall concentrate on $\{ C_{\ell}, \,\ell = 1,2,...,10 \}$. 

\subsection{The Hubble- and $q$-maps} \label{hubble-map_q-map}

This test involves the directional analyses of the Universe expansion through the mapping of
$H_0$ and $q_0$ in the celestial sphere. In other words, we adopt the opposite hemisphere method 
to accomplish this. Each pair of these hemispheres is well defined by the HEALpix pixelization 
scheme, such that we fit $H_0$ and $q_0$ according to the maximal likelihood technique (MLT) in 
each one of them, and thus constructing the Hubble-maps and $q$-maps, respectively. The MLT consists in
minimising the quantity

\begin{equation}
\label{eq:chi2}
\chi^2 = \sum_i\left(\frac{\mu_i-\mu_{\mathrm{th}}(z_i,\mathbf{p})}{\sigma_{\mu_i}}\right)^2 \;,
\end{equation}
where the set $(z_i, \mu_i, \sigma_{\mu_i})$ contains the observational information of the SNe data, i.e., redshift, distance moduli and associated uncertainty of the {\it{i-th}} object, respectively~\footnote{$\sigma_{\mu_i}$ is the error in the measurement of the SNe distance moduli, 
not to be confused with the discrete function $\sigma_j, \, j=1,\cdots,N_{\mbox{\tiny hems}}$, 
defined in section~\ref{sigma-map} to construct the sigma-map.} and $\mu_{\mathrm{th}}(z,\mathbf{p})$ is the distance modulus
\begin{equation}
\label{eq:mu_th}
\mu_{\mathrm{th}}(z,\mathbf{p}) =
5\log_{10}{ [ D_L(z,\mathbf{p}) ] } + 42.38 - 5\log_{10}(h),\,
\end{equation}
where $h \equiv H_0/100$, $H_0 \equiv 100 \,\mbox{Km} / \mbox{s} / \mbox{Mpc}$, and
$D_L(z,\mathbf{p})$ is the adimensional luminosity distance, whose arguments 
are the redshift $z$, in addition to the set of cosmological parameters $\mathbf{p}$ which describe the 
underlying cosmological model.

\subsection{Statistical significance tests} \label{MC-shuffle}

Once the Hubble- and $q$-maps have been constructed, it is of great relevance to test their statistical significance. Then, we follow two different approaches to accomplish this. In the first test one keeps the original set $(z, \mu, \sigma_{\mu})$ of each SNe and shuffles their galactic coordinates, as performed, 
e.g., by Kalus {\it et al.} (2013). This procedure will be called the {\it shuffle} test throughout this paper, whose objective is to check the dependence between each $(z_i, \mu_i, \sigma_{\mu_i})$ with the SNe angular distribution pattern in the sky. Assuming that a SN event can occur in any patch of the sky independently of its redshift, it is not expected a dramatic change in the hemispherical asymmetries unless the cosmic expansion and/or acceleration is really anisotropic, or the precision of the dataset is too limited. 

The second test also keeps the original $(z, \mu, \sigma_{\mu})$, however, the SNe positions are 
randomly chose so the SNe are isotropically redistributed on the celestial sphere. 
This test will be referred as {MC} from now on. 
In this case, we are able to verify the constraining potential of our anisotropy diagnostics in the case of 
idealistically isotropic datasets. If any anisotropic signal persists, it must be accounted to a possible 
violation of the CP, or the limitation of the observational sample. 

Finally, we should remark that, although MC and shuffle can be carried out for the Hubble and q-maps, 
only the former is actually suitable for the sigma-map. Since the sigma-map only requires
the angular position of the SNe events, the shuffle test is not convenient for these analyses 
as it maintains the original SNe distribution on the celestial sphere. Thus, as the sigma-map
is designed to quantify the uniformity of angular distributions, only the MC realisations
will be computed when the sigma-map test is performed.


\begin{table}
\begin{center}
\label{tab:tab_sigma_map} 
\begin{tabular}{ccc}
\hline
\hline 
Union2.1 sigma-map\\
\hline
\hline
dipole amplitude & $(l,b)$ \\
\hline
$+0.0017$ & $(295.00^{\circ}, \; -63.45^{\circ})$ \\
\hline
\hline
JLA sigma-map\\
\hline
\hline
dipole amplitude & $(l,b)$ \\
\hline
$+0.0020$ & $(195.00^{\circ}, \; -81.22^{\circ})$ \\
\hline
\hline
\end{tabular}
\end{center}
\caption{Amplitude and celestial position of the maximal sigma-map dipole contribution 
for the SNe Union2.1 (top), and the JLA datasets (bottom). 
Remember that this amplitude is given in arbitrary units, and that 
$l \in [0^{\circ},360^{\circ}], \, b \in [-90^{\circ},90^{\circ}]$; for instance, 
the North (South) galactic pole has coordinates $l$ arbitrary and $b=90^{\circ}$ 
($b=-90^{\circ}$). In all the cases, the error in the angular estimates is $\pm 3.66^{\circ}$.}
\end{table}

\section{Analyses and Results} \label{analyses-results}

\subsection{The SNe celestial maps} \label{SNe-map}


\begin{figure*}
\includegraphics[width = 6.0cm, height = 8.75cm, angle = +90]{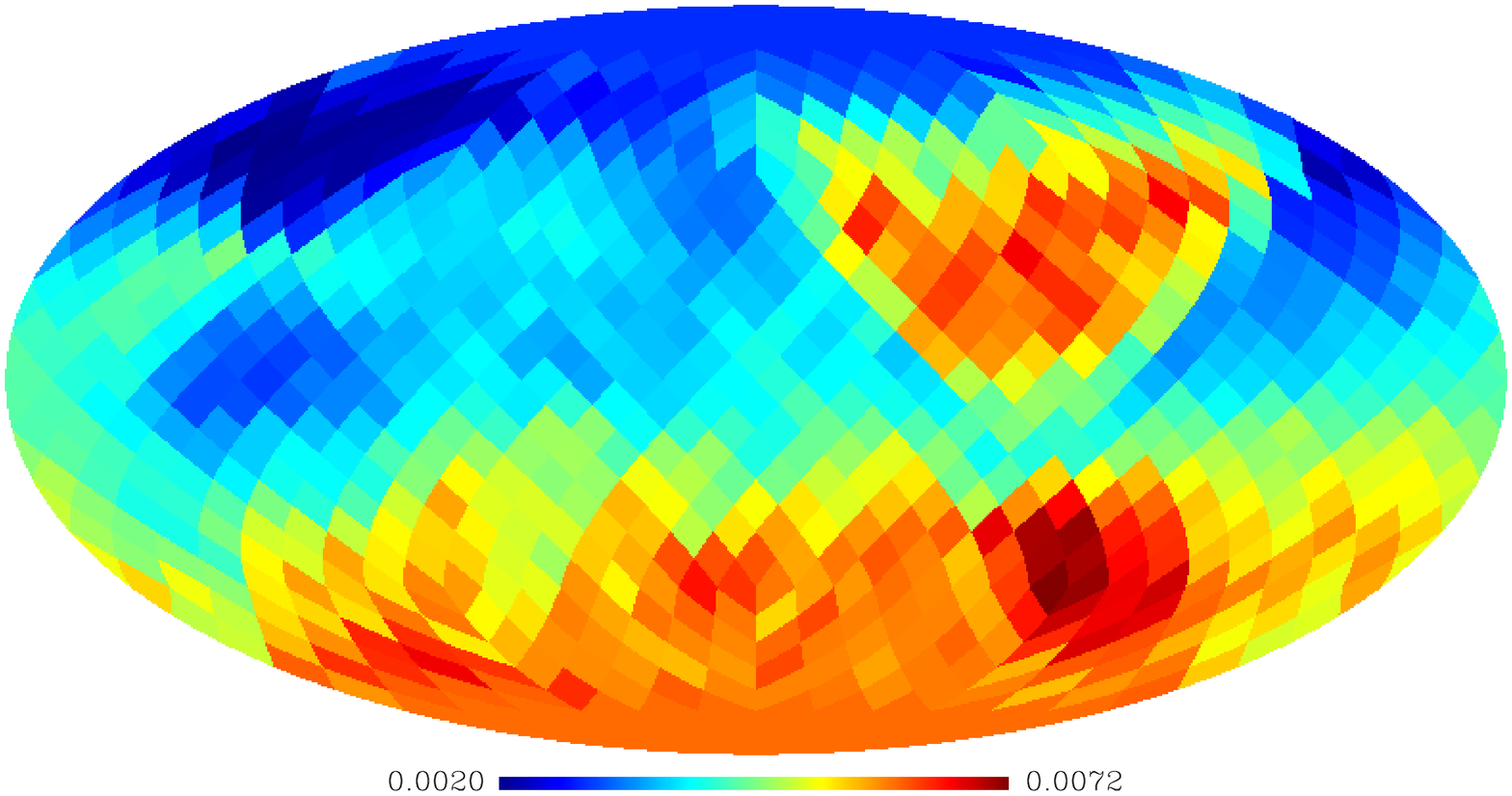}
\hspace{0.3cm}
\includegraphics[width = 6.0cm, height = 8.75cm, angle = +90]{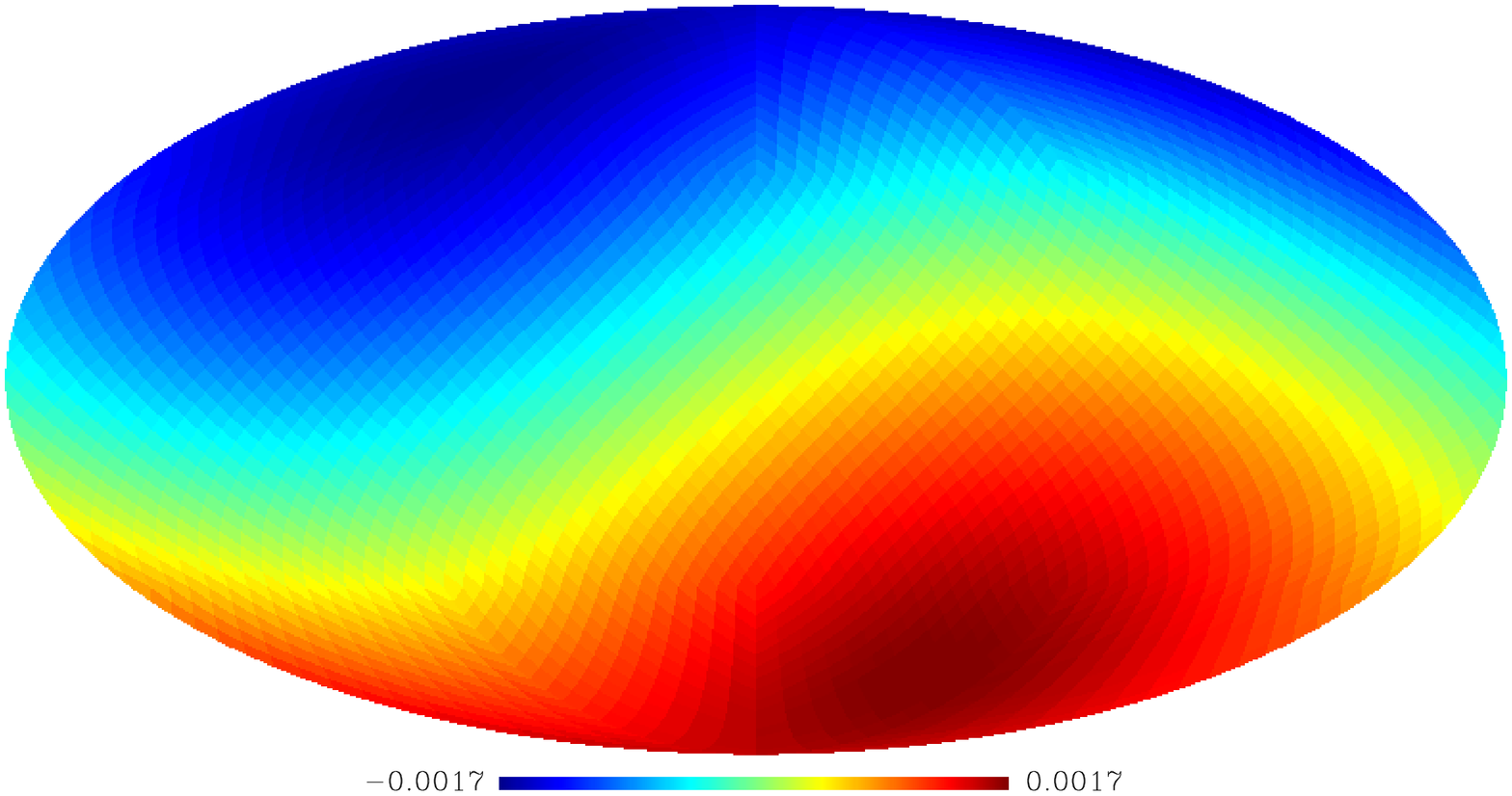}
\caption{{\it Left panel:} Sigma-map of the low-z, mask20 Union2.1 catalog, showing their angular correlation signatures. The units are arbitrary. 
{\it Right panel:} The dipole term, $\ell = 1$, of this sigma-map, which indicates the
direction in the sky where the mininal and maximal two-point angular correlations occur.}
\label{fig3}
\end{figure*}



\begin{figure*}
\includegraphics[width = 6.0cm, height = 8.75cm, angle = +90]{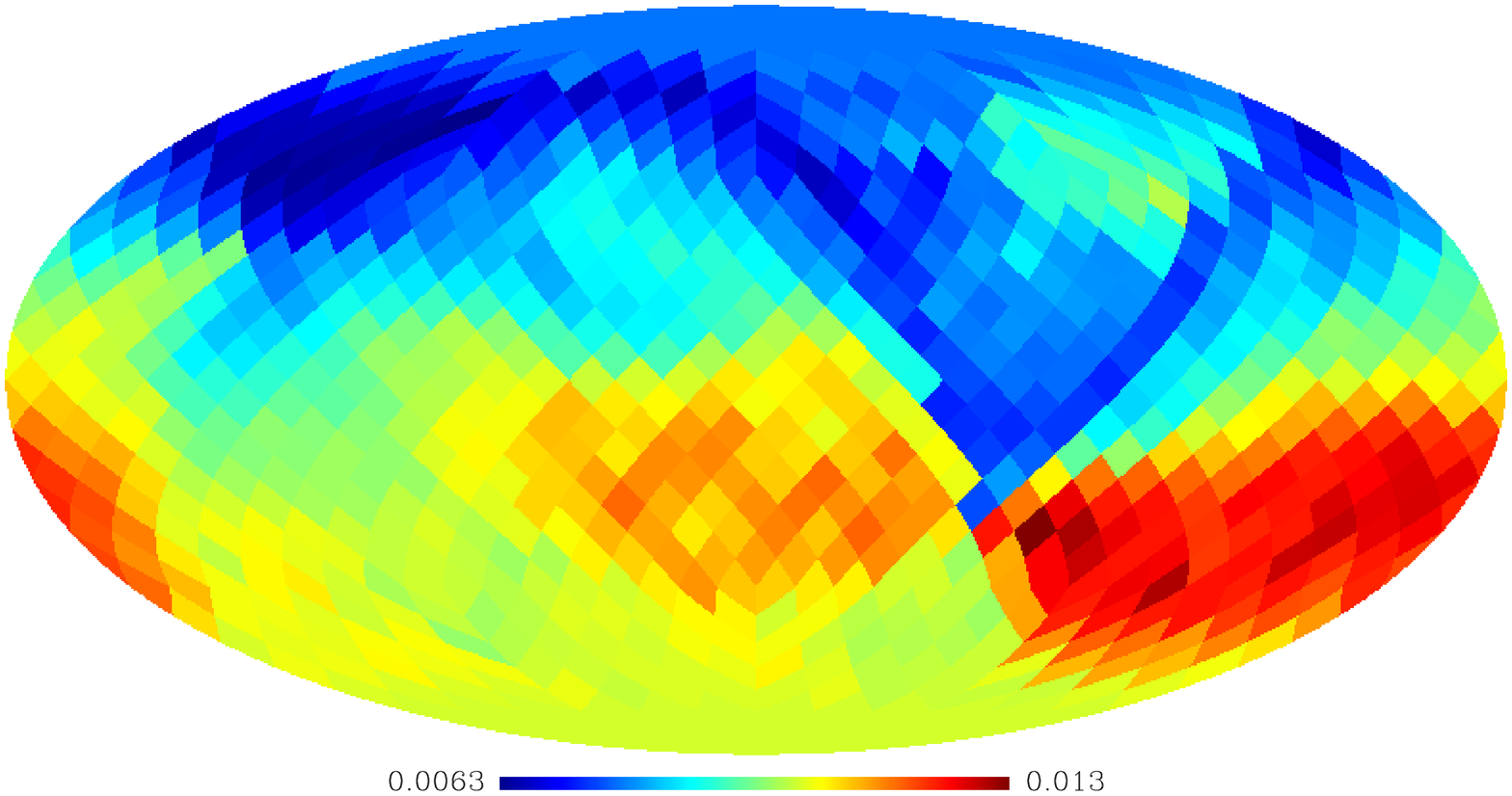}
\hspace{0.3cm}
\includegraphics[width = 6.0cm, height = 8.75cm, angle = +90]{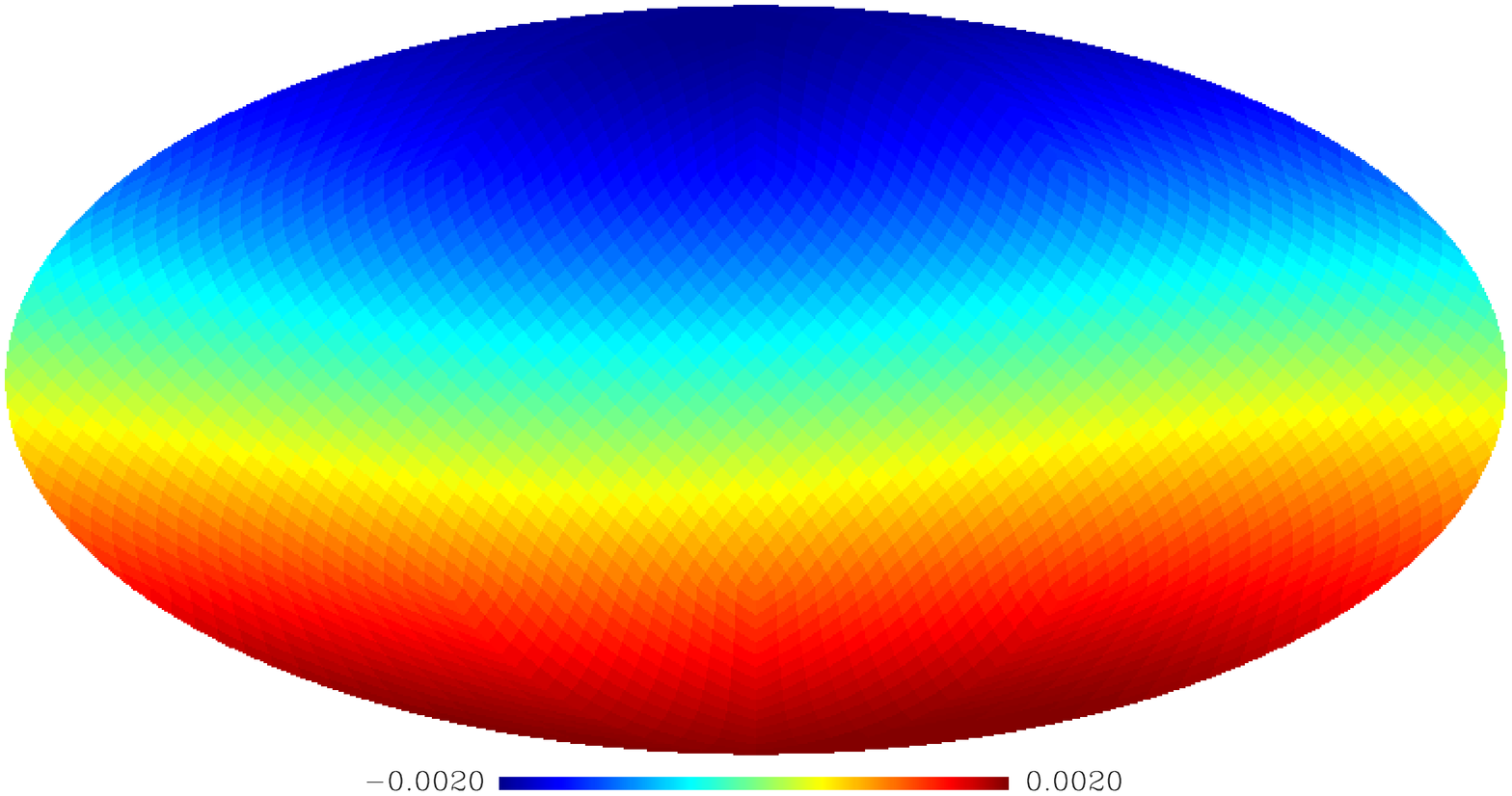}
\caption{{\it Left panel:} Sigma-map of the low-z, mask20 JLA catalog, showing their angular correlation signatures. The units are arbitrary. 
{\it Right panel:} The dipole term, $\ell = 1$, of this sigma-map, which indicates the
direction in the sky where the mininal and maximal two-point angular correlations occur.}
\label{fig4}
\end{figure*}


Considering only low-$z$ SNe ($z \leq 0.20$) we plot the maps of the SNe distribution in the celestial sphere in figures~\ref{fig1} and \ref{fig2} for the Union2.1 and JLA catalogs, respectively. In the top panel of both figures, it is possible to verify the more and less populated sky regions as well as a small number of SNe lying on the galactic plane due to the high thermal emission of dust in the Milky Way on that region. Hence, we also plot the SNe sky map with a galactic cut in the $|b| \leq 20^{\circ}$ plane, which can be visualised on the right panel of both figures. We name this galactic cut {\it mask20}, and it will be used in our analyses in order to avoid any biases in our results due to incomplete coverage in this strip of the sky.

It is worthy to remark that, even though the mask20 cuts one third of the whole celestial sphere area, there is 
not so much data there due to the high optical extinction on this galactic region. 
The full-sky Union2.1 sample, limited to $z \le 0.20$, contains 230 data points whereas the masked dataset has 211 events. It accounts less than 10\% of the whole sample, therefore, it should not considerably affect the results of our analyses. For the JLA sample, the same procedure diminishes the number of SNe from 317 to 302.


\begin{figure*}
\includegraphics[width = 6.0cm, height = 8.0cm, angle = -90]{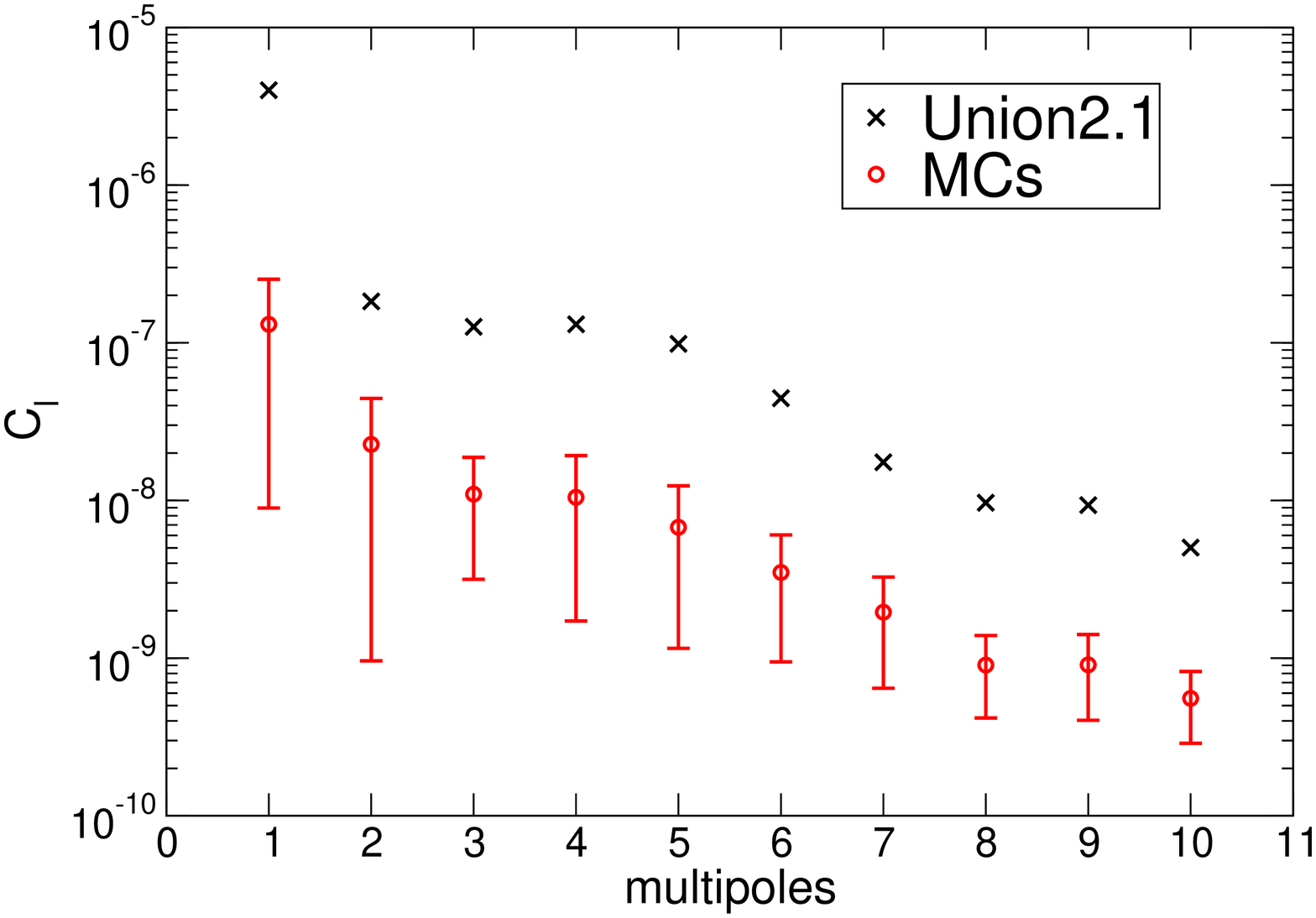}
\hspace{0.3cm}
\includegraphics[width = 6.0cm, height = 8.0cm, angle = -90]{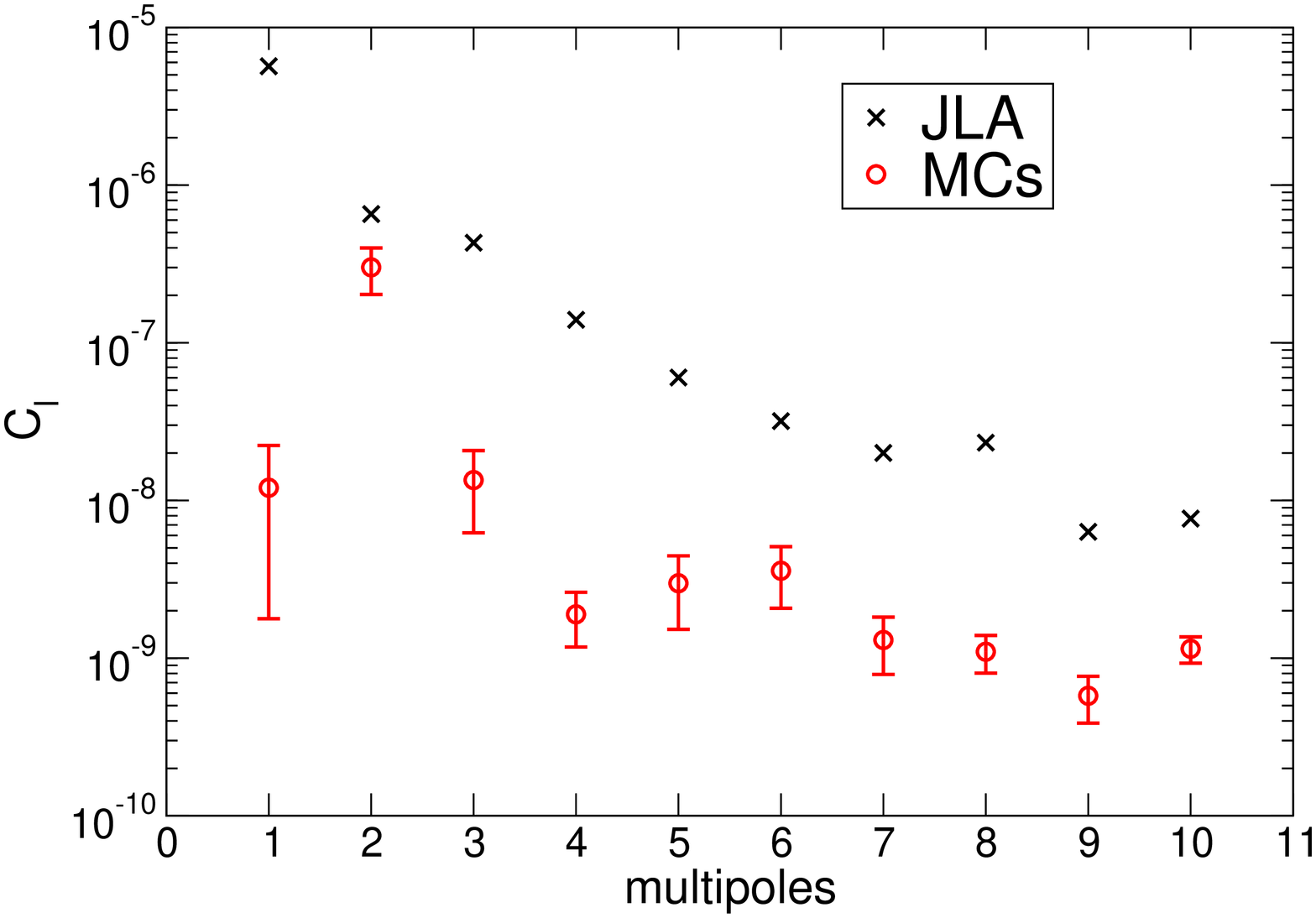}
\caption{{\it Left panel:} The angular power spectrum, $\{C_{\ell}\}$, up to $\ell = 10$, 
of the sigma-map for the Union2.1 catalog. {\it Right panel:} the same for the JLA catalog. 
The crosses represent the values for the original datasets, while the red circles
assign the average spectra from sigma-maps obtained from 500 MC realisations. Their 
error bars are estimated using the median absolute deviation of each coefficient of these spectra.
The $\chi^2$ values, for 9 degrees of freedom, between real and simulated data
are $5.6 \times 10^3$ and $3.5 \times 10^5$ for the Union2.1 and JLA samples, 
respectively, indicating that both SNe datasets are anisotropic at a high statistically significant level.}
\label{fig5}
\end{figure*}


\subsection{The sigma-maps results} \label{sigma-map-results}

The results for the sigma-map adopting the $z \leq 0.20$ range and the mask20 galactic cut are presented in figures~\ref{fig3} (Union2.1) and \ref{fig4} (JLA).
The left panels of both figures show the original sigma-map, while the right panels feature their respective dipole-only contributions. Table 1 displays the amplitude of this dipole, in addition to the celestial position where it occurs in galactic coordinates. It is worthy remarking that the choice of $N_{\mathrm side} = 8$ has been made for our plots so that the celestial sphere has 768 centres of hemispheres. 

From these analyses, we can readily note that both datasets are not perfectly isotropic, specially in the JLA sample, as the 2PACF yields significantly higher numbers in some regions of the sky than others, pointing out a potential observational bias for anisotropic signal due to sky patches with under-sampled and over sampled SNe data. We measure the degree of anisotropic distribution of the SNe data by comparing its angular power spectrum with the mean of 500 sigma-maps power spectra produced by MC realisations, that is the case where the data points are isotropically distribuited in the sky. These results are shown in the figure~\ref{fig5} for both SNe catalogs. As the galactic cut encompasses one third of the whole celestial sphere, these simulated datasets contain 33\% additional events compared to the original sample, which would be the roughly expected number of observed SNe if the extinction around the galactic plane was much less intense. 
 
The $\chi^2$ goodness-of-fit test between the SNe and these simulated data gives us a measure of the consistency or deviation from the statistical isotropy hypothesis (that is, the null hypothesis) of the SNe dataset. We obtain $\chi^2 = 5.6 \times 10^3$ and $\chi^2 = 3.5 \times 10^5$, respectively, for 9 degrees of freedom, indicating that the SNe data from both catalogs are anisotropic at a high confidence level. Moreover, this multipole expansion analyses reveal that the dipole terms are much higher than the higher $\ell$ order, thus showing the significance of their contributions.


\begin{table}
\begin{center}
\label{tab:tab_hubble_map} 
\begin{tabular}{ccc}
\hline
\hline
Union2.1 Hubble-map \\
\hline 
dipole amplitude & $(l,b)$ \\
\hline
$+0.015$ & $(326.25^{\circ}, \; 12.02^{\circ})$ \\
\hline
\hline
JLA Hubble-map \\
\hline 
dipole amplitude & $(l,b)$ \\
\hline
$+0.025$ & $(58.00^{\circ}, \; -60.43^{\circ})$ \\
\hline
\hline
\end{tabular}
\end{center}
\caption{Amplitude and celestial position of the maximal Hubble-map dipole contribution 
for the SNe Union2.1 (top), and the JLA datasets (bottom). 
Remember that this amplitude is given in arbitrary units. 
In all the cases, the error in the angular estimates is $\pm 3.66^{\circ}$.}
\end{table}



\begin{table}
\begin{center}
\label{tab:tab_q_map} 
\begin{tabular}{ccc}
\hline
\hline
Union2.1 q-map\\
\hline
dipole amplitude & $(l,b)$ \\
\hline
$+0.44$ & $(174.38^{\circ}, \; 27.28^{\circ})$ \\
\hline
\hline
JLA q-map\\
\hline
dipole amplitude & $(l,b)$ \\
\hline
$+0.54$ & $(45.00^{\circ}, \; -51.26^{\circ})$ \\
\hline
\hline
\end{tabular}
\end{center}
\caption{Amplitude and celestial position of the maximal q-map dipole contribution 
for the SNe Union2.1 (top), and the JLA datasets (bottom). 
Remember that this amplitude is given in arbitrary units. 
In all the cases, the error in the angular estimates is $\pm 3.66^{\circ}$.}
\end{table}


\vspace{2.0cm}

\subsection{The hubble and $q$-maps results} \label{hubble-q-map}

As described in section 2, we construct our maps by associating a pixel value with the best-fit of the parameter of interest in each one of the 768 hemispheres. Then, we are able to determine which sky patch presents the highest and lowest expansion rate (in case of Hubble-map), the same for the cosmological acceleration ($q_0$ case). As our analyses are performed in the $z \le 0.20$ range, we can use the cosmographic expansion to compute cosmological distances, which means that we do not need to assume any model for the dynamics of the Universe. Hence, the luminosity distance of the Eq.~(\ref{eq:mu_th}) is written as
\begin{equation}
\label{eq:DL_cosm}
D_L(y,\mathbf{p}) = y + \left(\frac{3 - q_0}{2}\right)y^2 \;; \quad y \equiv \frac{z}{1+z} \;,
\end{equation}
where $q_0 \equiv -\ddot{a}_0/(a_0 H_0^2)$. 

We have truncated our expansion series in the second-order term similarly to the procedure of Kalus {\it et al.} (2013), as the redshift range is short enough that the inclusion of third or higher order terms do not yield unambiguous results for $D_L$, independently of the value for $q_0$. It is also worthy noting that, as we have a two dimensional parametric space $(H_0, q_0)$, we marginalise over one of these parameters in order to map the other one on the celestial sphere. Therefore, marginalising over $q_0$ corresponds to the Hubble-map case while marginalising over $H_0$ gives us the $q$-map.


\begin{figure*}
\includegraphics[width = 6.0cm, height = 8.75cm, angle = +90]%
{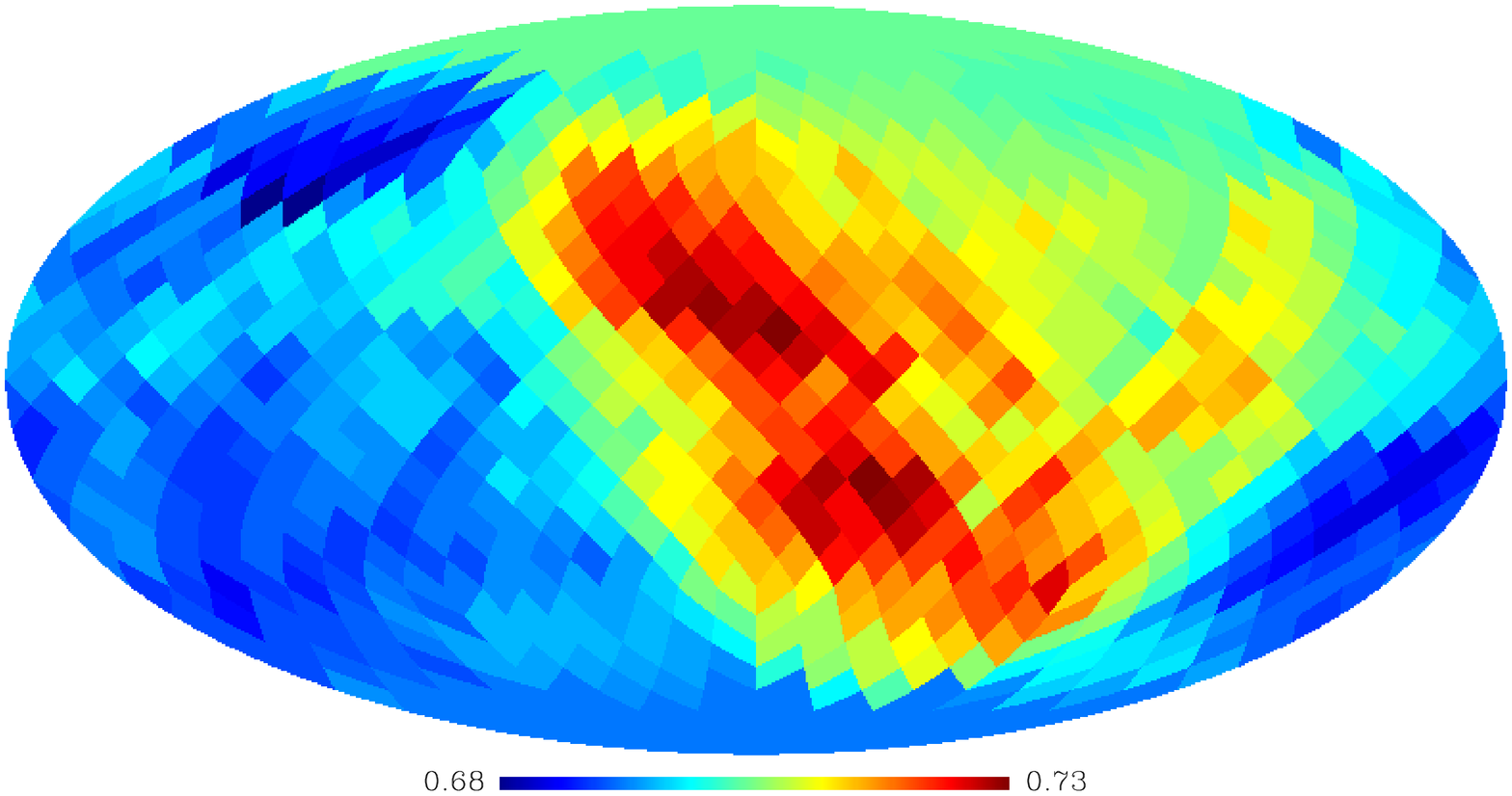}
\hspace{0.3cm}
\includegraphics[width = 6.0cm, height = 8.75cm, angle = +90]%
{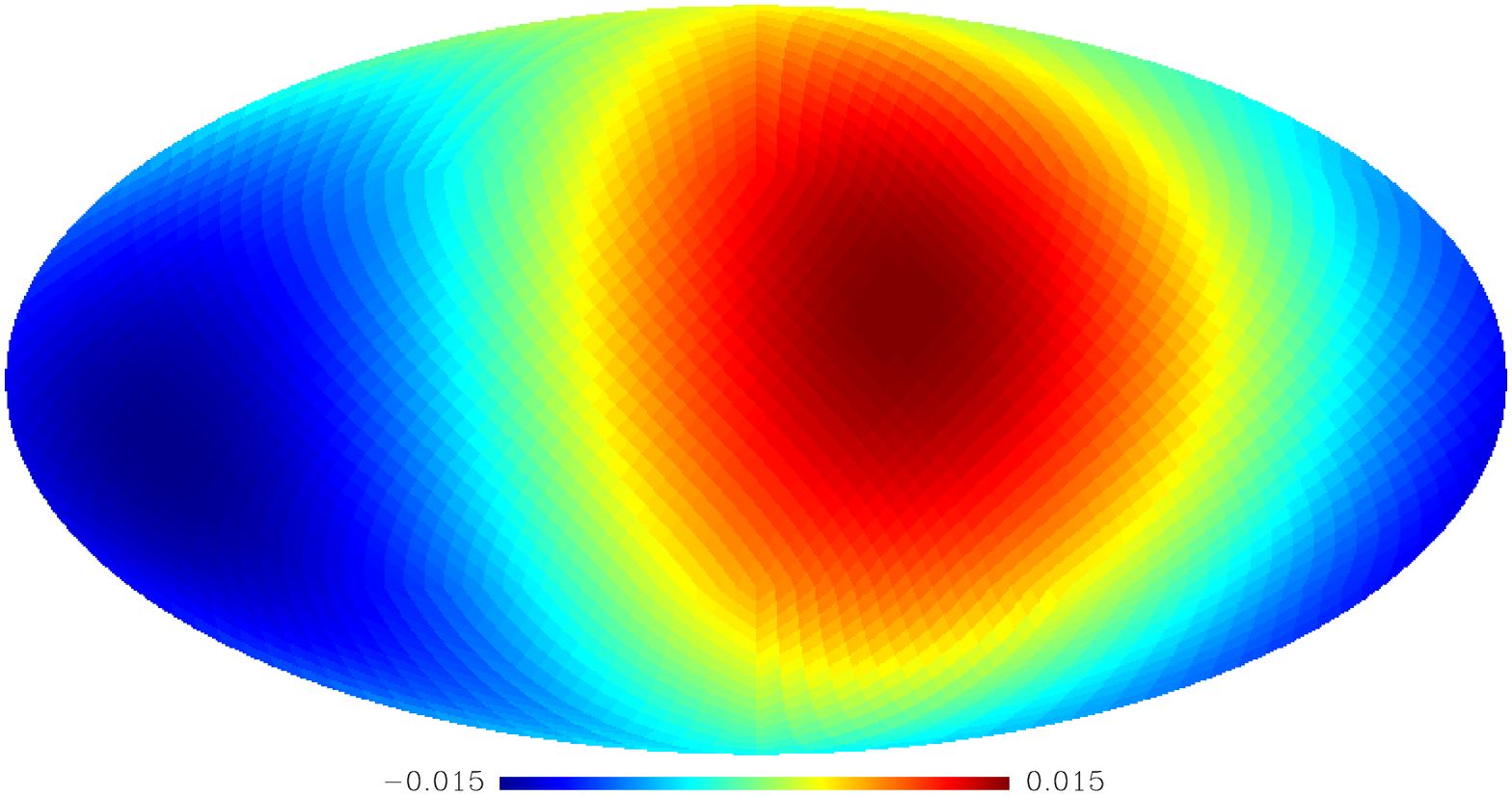}
\caption{{\it Left panel}: Hubble-map, in units of $h \equiv H_0/100$, for the low-z, mask20 Union2.1 
subset, showing the cosmological expansion in the celestial sphere. The lowest and highest values 
obtained for all 768 hemispheres are $h = 0.683$ and $h = 0.729$, respectively. 
{\it Right panel:} The dipole-only contribution term of the sigma-map, which indicates the
direction in the sky where the mininal and maximal values of $h$ occur.
}
\label{fig6}
\end{figure*}



\begin{figure*}
\includegraphics[width = 6.0cm, height = 8.75cm, angle = +90]%
{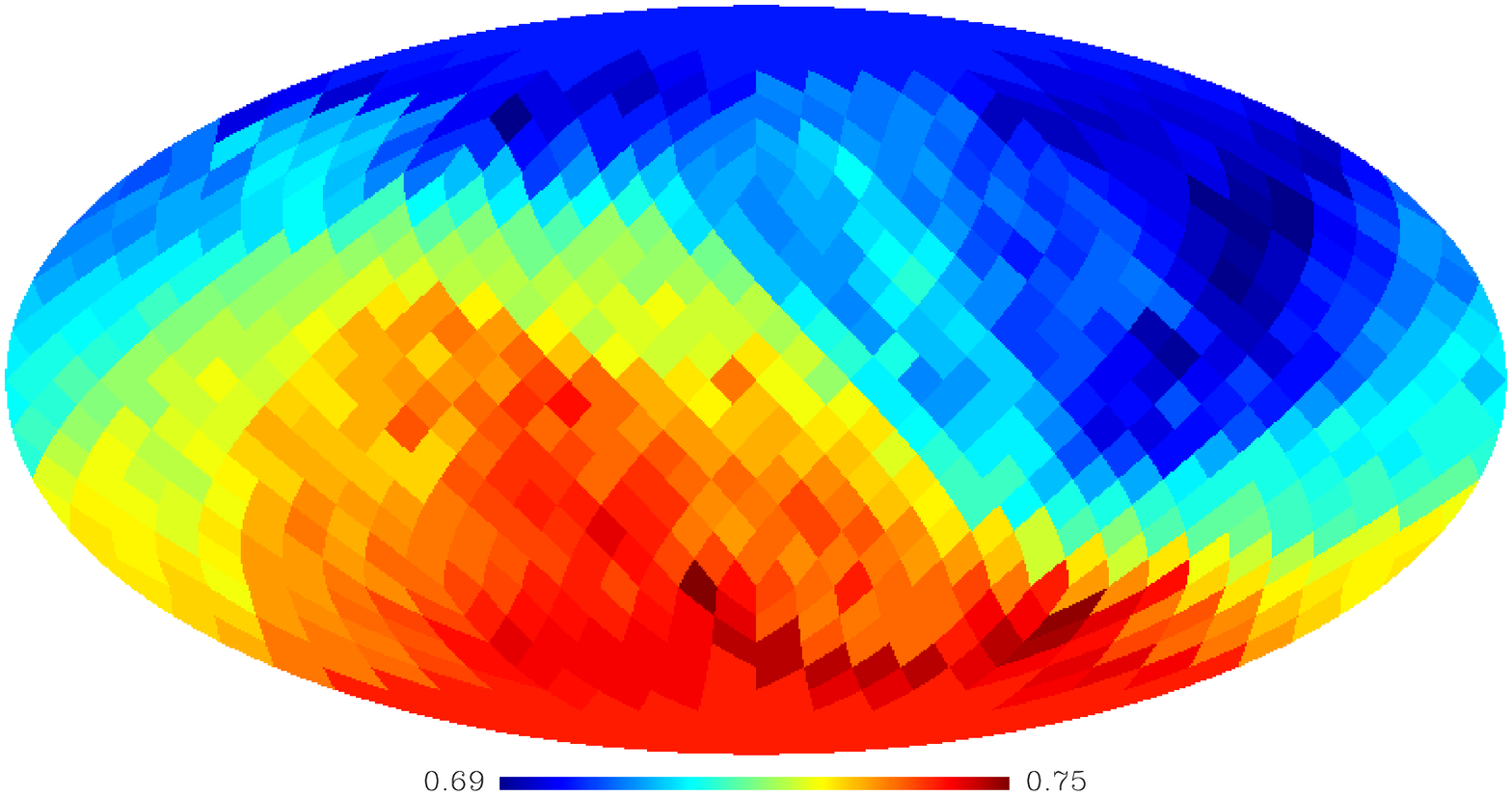}
\hspace{0.3cm}
\includegraphics[width = 6.0cm, height = 8.75cm, angle = +90]%
{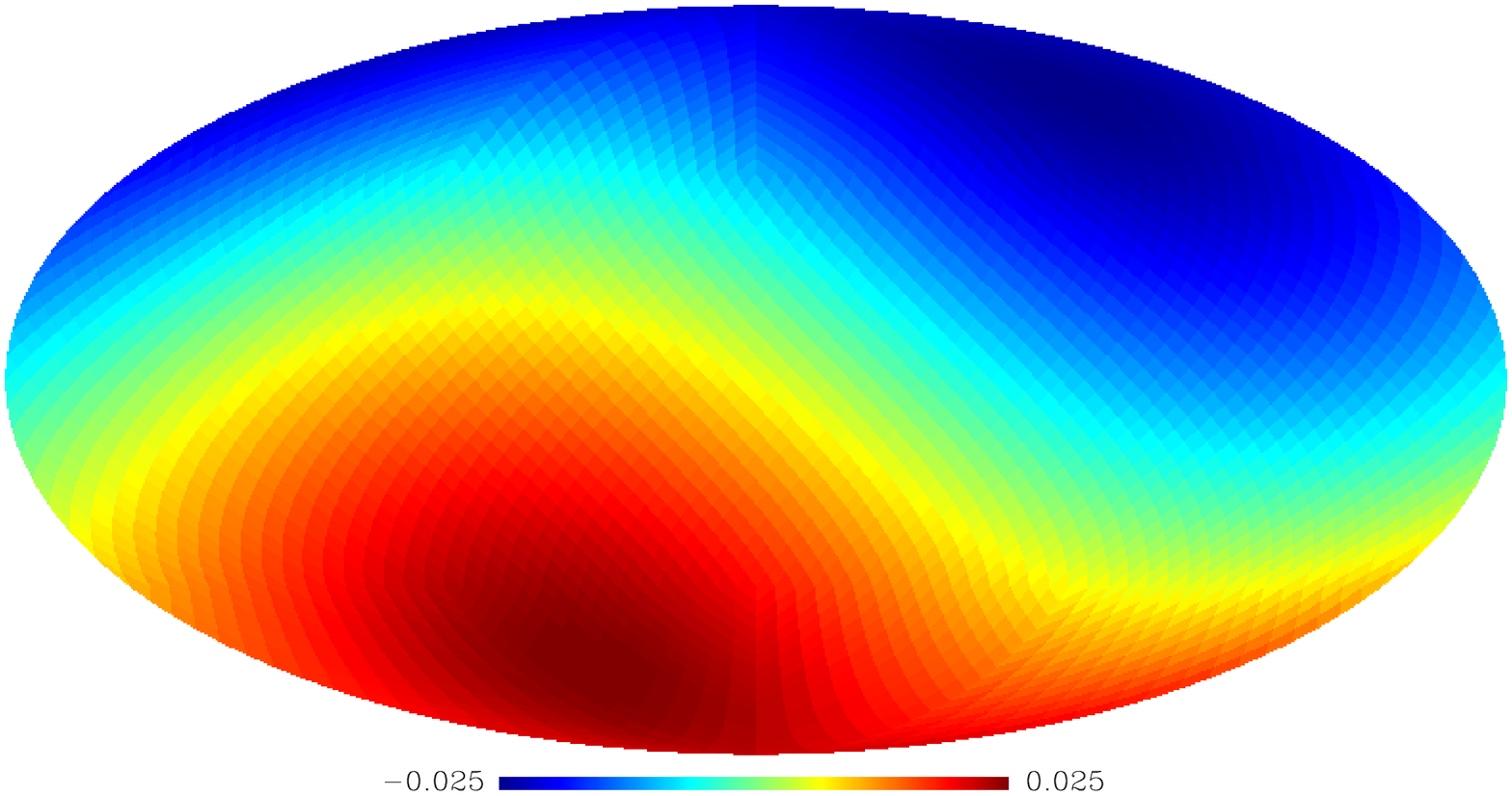}
\caption{ {\it Left panel}: Hubble-map, in units of $h \equiv H_0/100$, for the low-z, mask20 JLA subset, 
showing the cosmological expansion in the celestial sphere. The lowest and highest values obtained for 
all 768 hemispheres are $h = 0.695$ and $h = 0.750$, respectively. 
{\it Right panel:} The dipole-only contribution term of the sigma-map, which indicates the
direction in the sky where the mininal and maximal values of $h$ occur.
}
\label{fig7}
\end{figure*}

\begin{figure*}
\includegraphics[width = 6.0cm, height = 8.75cm, angle = +90]
{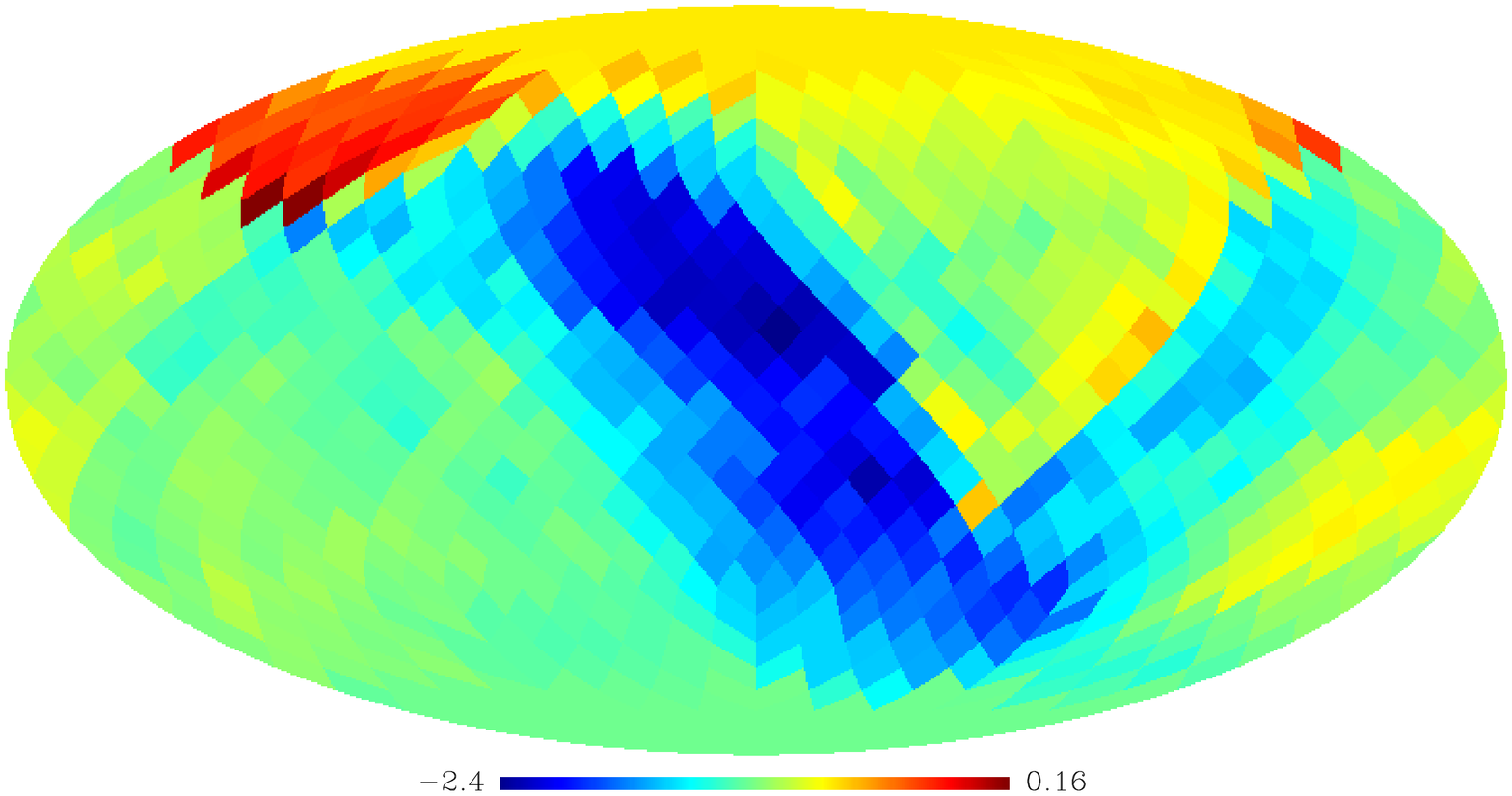}
\hspace{0.3cm}
\includegraphics[width = 6.0cm, height = 8.75cm, angle = +90]
{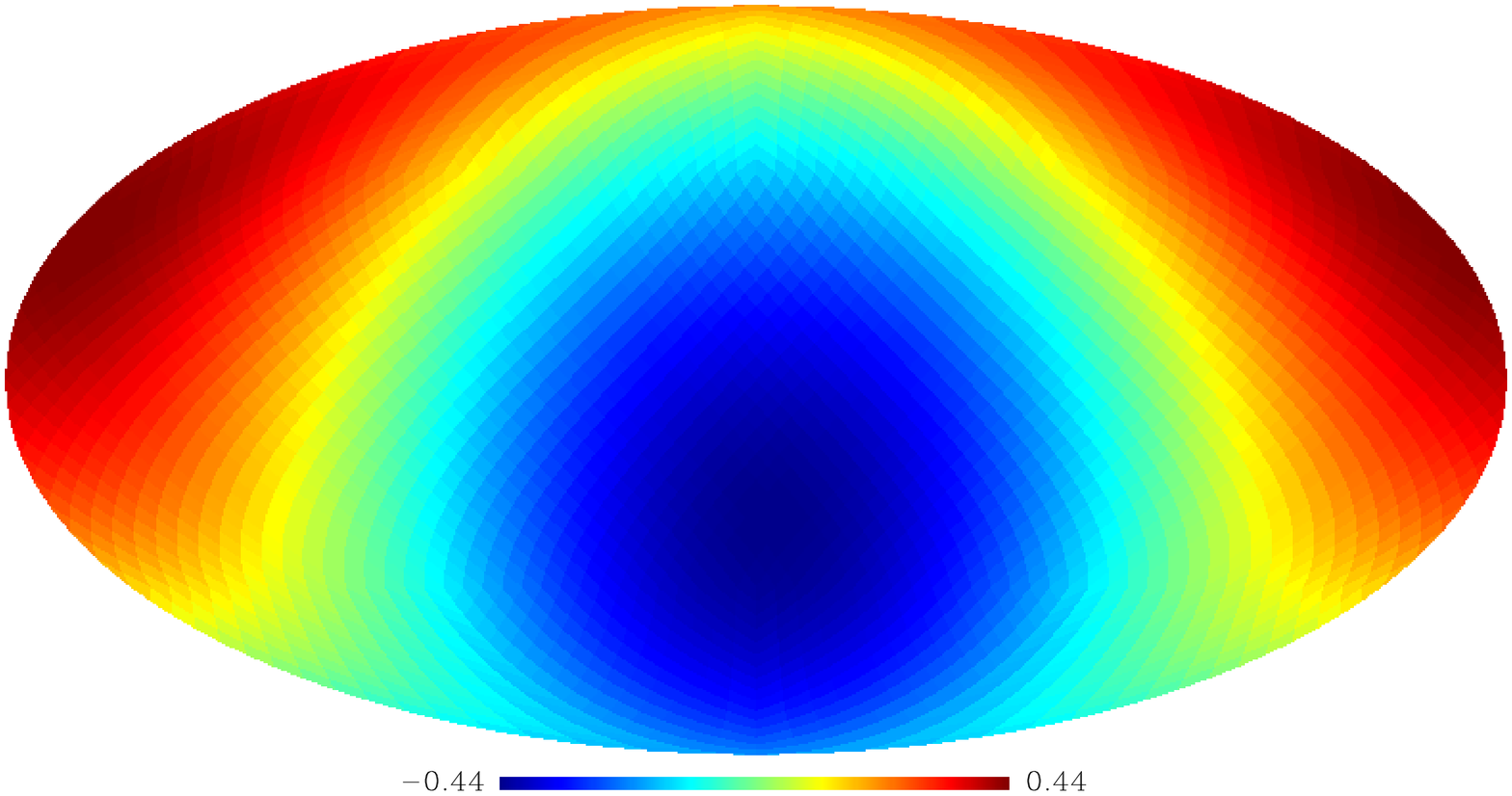}
\caption{{\it Left panel}: The q-map for the low-z, mask20 Union2.1 subset, showing the cosmological expansion in the celestial sphere. The lowest and highest values obtained for all 768 hemispheres are 
$q_0 = -2.40$ and $q_0 = +0.16$, respectively. 
{\it Right panel:} The dipole-only contribution term of the $q$-map, which indicates the
direction in the sky where the minimal and maximal values of $q_0$ occur. 
}
\label{fig8}
\end{figure*}
\begin{figure*}
\includegraphics[width = 6.0cm, height = 8.75cm, angle = +90]
{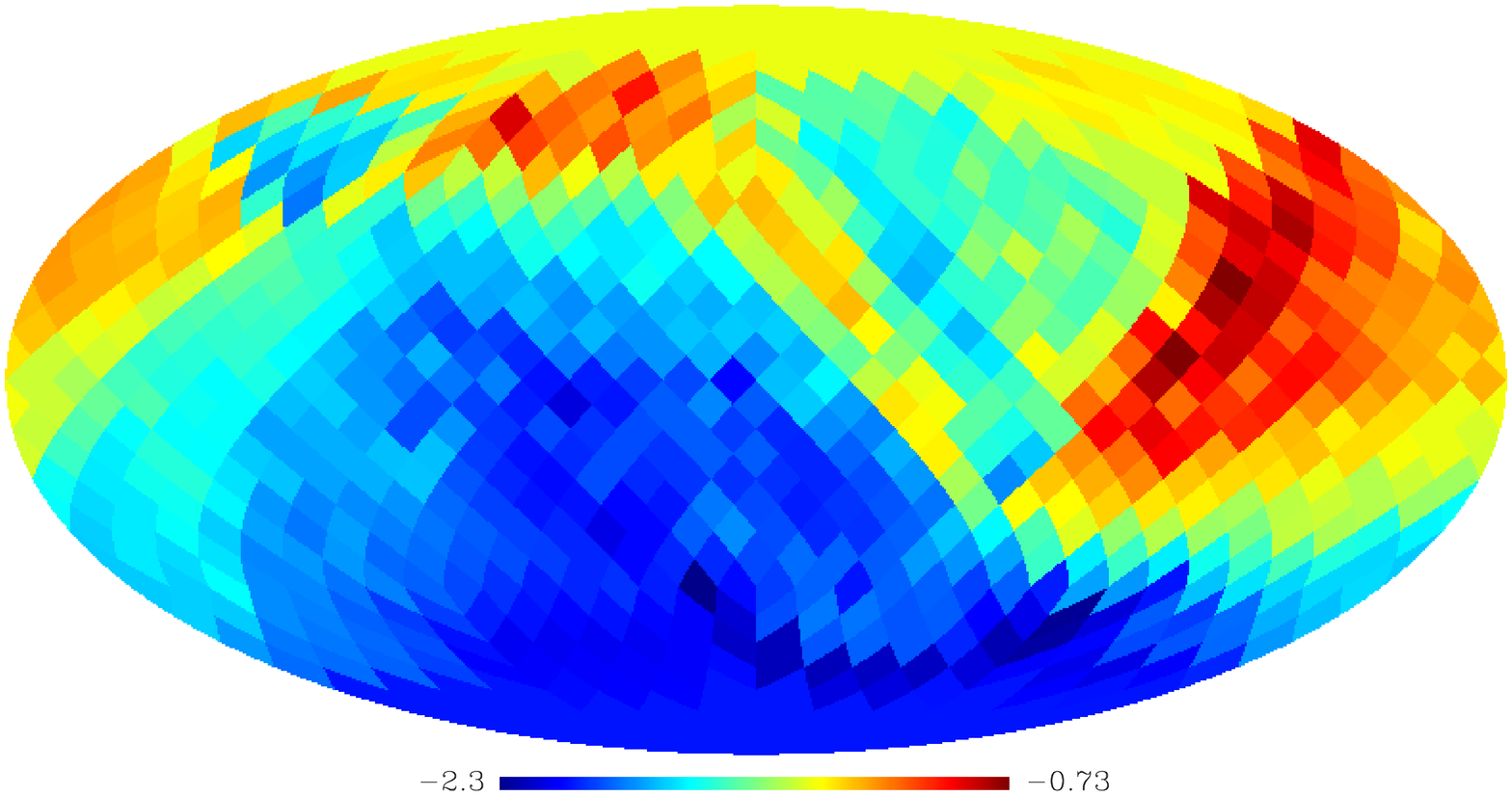}
\hspace{0.3cm}
\includegraphics[width = 6.0cm, height = 8.75cm, angle = +90]
{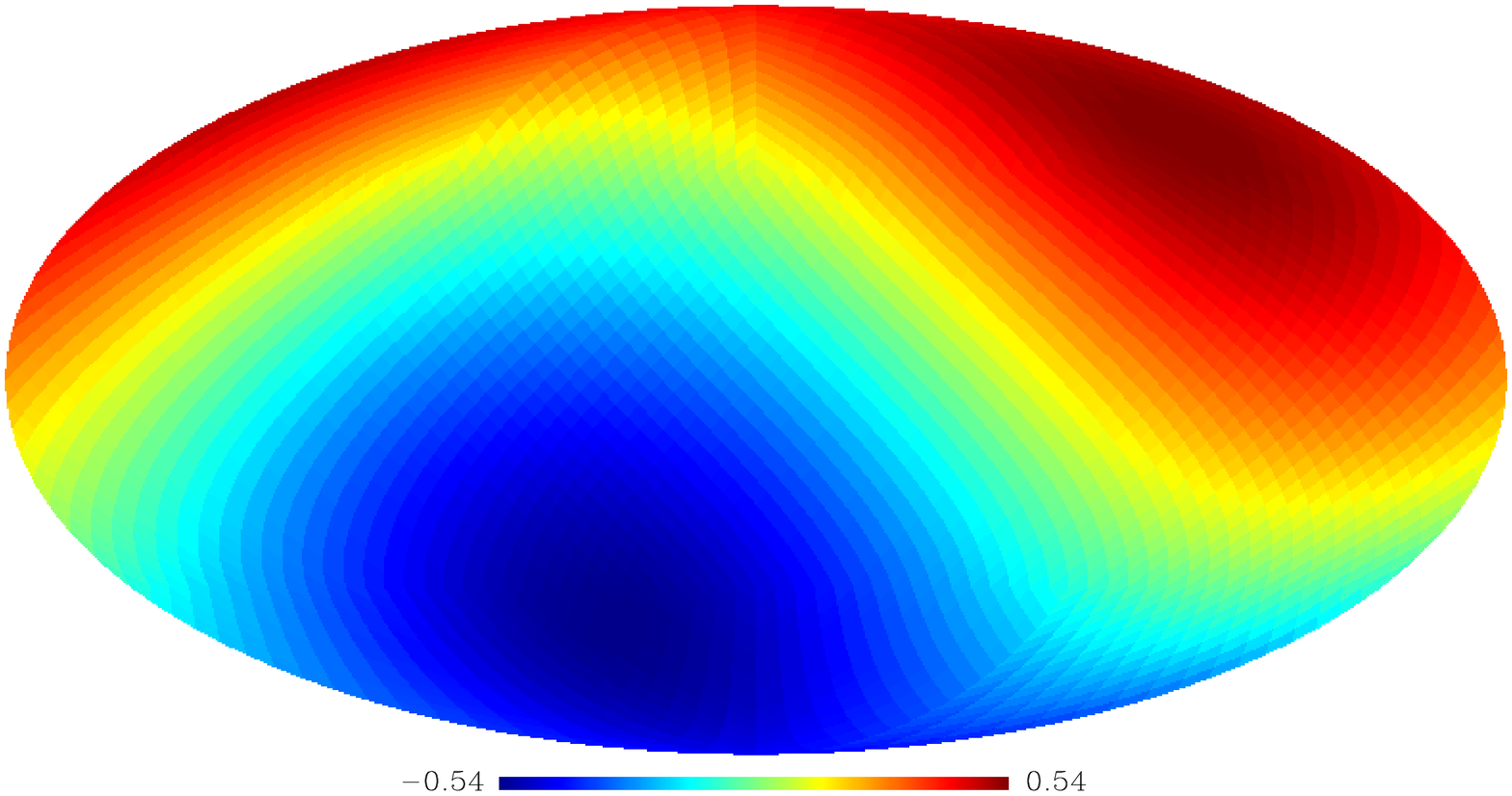}
\caption{{\it Left panel}: The q-map for the low-z, mask20 JLA subset, showing the cosmological 
expansion in the celestial sphere. The lowest and highest values obtained for all 768 hemispheres are 
$q_0 = -2.35$ and $q_0 = -0.73$, respectively. 
{\it Right panel:} The dipole-only contribution term of the $q$-map, which indicates the
direction in the sky where the mininum and maximal values of $q_0$ occur. 
}
\label{fig9}
\end{figure*}

The results for the Hubble-maps are shown in figures~\ref{fig6} and \ref{fig7} for the Union2.1 and JLA samples, respectively, whereas the $q$-maps are displayed in figures~\ref{fig8} (Union2.1 case) and~\ref{fig9} (JLA). We found the maximal asymmetry in the Hubble-map of $\delta H_0 \equiv {H_0}_{\rm max} - {H_0}_{\rm min} = 4.6$ for the Union2.1 and $\delta H_0 = 5.6$ for the JLA case. This is larger than the measurement of asymmetry reported by Kalus {\it et al.} (2013), i.e., $\delta H_0 = 2.6$, who, however, 
used a smaller SNe sample (183 objects) and fixed the value of the decelaration parameter at $q_0 = -0.601$ in their analyses. The $q$-maps analyses yield $\delta q_0 = 2.56$ and $\delta q_0 = 1.62$ for the Union2.1 and JLA SNe samples, respectively.  
 
The left panels of these figures present the dipole-only contribution of each of these maps. These analyses allow us to compute the preferred axes of our anisotropy diagnostics, that is $H_0$ and $q_0$. In the Hubble-map case, for instance, we obtained the highest expansion rate of the Universe towards $(l,b) = (326.25^{\circ},12.02^{\circ})$ for the Union2.1 catalog and $(l,b) = (58.00^{\circ}, -60.43^{\circ})$ for the JLA sample, as described in Table 2. 
For the $q$-map case, shown in Table 3, the preferred direction for the cosmological acceleration is $(l,b) = (354.38^{\circ},-27.28^{\circ})$ (Union2.1) 
and $(l,b) = (225.00^{\circ},-51.26^{\circ})$ (JLA). Even though there are some interception between both samples, our $H_0$ and $q_0$ analyses give rather different results for the amplitude of both anisotropy diagnostics as well as the directions in the sky where they occur. 

For completeness, we also compare our results with the preferred axes reported from different analyses (using the Union SNe compilation) in Table 4. Finally, we  also stress that this results seem to be consistent with the bulk-flow motion measurements reported by Turnbull {\it et al.} (2012), who detected a bulk-flow of $249 \pm 76$ km/s towards $(l,b) = (319^{\circ} \pm 18^{\circ}, 7^{\circ} \pm 14^{\circ})$ using nearby SNe, as well as Colin {\it et al.} (2011), whose bulk-flow of $(l,b) = (299^{\circ}, 13^{\circ})$ was computed using the Union2 catalog at $0.015 < z < 0.2$, the same redshift regime that we perform our analyses, and also Feindt {\it et al.} (2013), who obtained the result of $v_{bf} = 292 \pm 96$km/s towards $(l,b) = (290^{\circ} \pm 22^{\circ}, 15^{\circ} \pm 18^{\circ})$ with the most nearby SNe ($0.015 < z < 0.035$) from this same sample. Thus, it is possible to identify the anisotropic signal of the Union2.1 maps with this phenomenon. In the JLA case, however, we can not compare our 
results with previous reports since this sample contains many SNe events which are not present in the Union2.1 dataset.

We also performed a simple test whose goal is to verify whether the goodness of these fits coincide with the minimal and maximal values of $H_0$ and $q_0$ that we have discussed earlier. Naturally, reduced $\chi^2$, i.e., $\chi^2_\nu = \chi^2/\nu$, is adopted for these tests, where $\nu$ represents the number of degrees of freedom in each hemisphere. 
We note that the poorest fits for each parameter 
range from $\sim 0.6$ to $1.3$ in the Union2.1 maps and $\sim 0.45$ to $0.75$ in the JLA case, where these values do not coincide with the minimal and maximal $H_0$ or $q_0$ in any case. Thus, we conclude that the MLT we have carried out do not introduce a bias on the coslogical expansion (or acceleration) signals. Similar analyses have also been performed by Kalus {\it et al.} (2013), where no impact between their Hubble-maps and their $\chi^2$-maps have been detected as well.

\subsection{The correlation analyses}


\begin{table}
\begin{center}
\label{tab:tab_max_anis} 
\begin{tabular}{ccc}
\hline
\hline
$(l,b)$ & Ref.\\
\hline
\hline
$({309^{\circ}}^{+23^{\circ}}_{-03^{\circ}}, {18^{\circ}}^{+11^{\circ}}_{-10^{\circ}})$ & Antoniou \& Perivolaropoulos, 2010 \\
$({314^{\circ}}^{+20^{\circ}}_{-13^{\circ}}, {28^{\circ}}^{+11^{\circ}}_{-33^{\circ}})$ & Cai \& Tuo, 2012 \\
$(325^{\circ}, -19^{\circ})$ & Kalus {\it et al.}, 2013 \\
$(187^{\circ}, 108^{\circ})$ & Zhao {\it et al.}, 2013 \\
$(306^{\circ}, -13^{\circ})$ & Cai {\it et al.}, 2013 \\
$({307.1^{\circ}}^{+16.2^{\circ}}_{-16.2^{\circ}}, {-14.3^{\circ}}^{+10.1^{\circ}}_{-10.1^{\circ}})$ & Yang {\it et al.}, 2014 \\
$({306.1^{\circ}}^{+18.7^{\circ}}_{-18.7^{\circ}}, {-18.2^{\circ}}^{+11.2^{\circ}}_{-11.2^{\circ}})$ & Chang {\it et al.}, 2014 \\
$({309.2^{\circ}}^{+15.8^{\circ}}_{-15.8^{\circ}}, {-8.6^{\circ}}^{+10.5^{\circ}}_{-10.5^{\circ}})$ & Wang {\it et al.}, 2014 \\
\hline
$(326.25^{\circ}, 12.02^{\circ})$ & this paper (hubble-map, Union2.1) \\
$(58.00^{\circ}, -60.43^{\circ})$ & this paper (hubble-map, JLA) \\
$(354.38^{\circ}, 27.28^{\circ})$ & this paper (q-map, Union2.1) \\
$(225.00^{\circ}, 51.26^{\circ})$ & this paper (q-map, JLA) \\
\hline
\hline
\end{tabular}
\end{center}
\caption{ The axes reported in the literature, and our results, along which the maximal anisotropy occurs.} 
\end{table}


In Table 5 we display the correlation between each map, where the Pearson's coefficient $\rho$ 
have been adopted as a diagnostics for this purpose. 
For the Union2.1 dataset, it is possible to note that the Hubble-map and $q$-map show a significant correlation: 
$\rho \simeq -0.70$, roughly meaning that in the sky regions where $H_0$ is large $q$ is low 
and negative, as expected by the definition of $q$. 
Yet, the most important cases are the correlation between Hubble-maps and $q$-maps versus 
the sigma-maps. Despite the visual correlation between the celestial regions 
containing a large number of SNe 
(Fig.~\ref{fig1}) and the regions where $H_0$ achieves its lowest values (Fig.~\ref{fig4}), hence 
indicating a significant observational bias in our results as a matter of fact there is a small 
correlation between the Hubble-map as compared with the sigma-map, with Pearson's 
coefficient $\rho \simeq +0.059$. This means that over-sampled regions do not necessarily  
influence the Hubble diagram result. 
Similarly, there is weak anti-correlation, $\rho \simeq -0.200$ between the $q$-map and the 
sigma-map. These analyses let us to realise an overall assessment that the anisotropic SNe 
distribution does not 
imply an {\it a priori} influence in the anisotropic Hubble- and $q$-maps results. 
Finally, we have also verified if there is a correlation between Hubble- and $q$-maps and their 
corresponding $\chi^2$-map, finding out that this correlation is small in both cases. 
This result, therefore, confirms 
the validity of our approach regarding the absence of bias from 
the statistical technique we have adopted to map the cosmological parameters in the sky.  


Nevertheless, these analyses provide distinct results for the JLA compilation, which 
can also be checked on table 5. 
There is a relevant correlation between Hubble and $q$-maps with the sigma-maps, evaluated on 
$\rho = +0.65$ and $\rho = -0.45$,
respectively, besides a stronger correlation occurs between the Hubble and $q$-maps: $\rho = -0.91$. 
These results indicate that the incomplete sky coverage of this sample indeed affects 
the anisotropy of the cosmic expansion and acceleration. 

Finally, we have also computed the correlation between the maps obtained for
each dataset. Although there is a suggestive correlation between the sigma-maps,
this trend is not followed by the Hubble- and $q$-maps, which are just mildly correlated. 
This analysis reflects the results we have discussed before, where we have shown 
that the axes of maximal anisotropic expansion (and acceleration) are very different
for each catalog.


\begin{table}
\begin{center}
\label{tab:tab_corr} 
\begin{tabular}{ccc}
\hline
\hline
Union2.1 maps & $\rho$ \\
\hline
Hubble-map vs. sigma-map & $+0.059$  \\ 
q-map vs. sigma-map & $-0.200$   \\ 
q-map vs. Hubble-map & $-0.702$  \\  
Hubble-map vs. ${\chi^2_\nu}_{\rm Hubble-map}$ & $+0.054$  \\ 
q-map vs. ${\chi^2_\nu}_{\rm q-map}$ & $-0.391$  \\ 
\hline
\hline
JLA maps & $\rho$ \\
\hline
Hubble-map vs. sigma-map & $+0.651$  \\ 
q-map vs. sigma-map & $-0.446$   \\ 
q-map vs. Hubble-map & $-0.915$  \\  
Hubble-map vs. ${\chi^2_\nu}_{\rm Hubble-map}$ & $-0.130$  \\ 
q-map vs. ${\chi^2_\nu}_{\rm q-map}$ & $-0.346$  \\ 
\hline
\hline
Union2.1 vs JLA maps & $\rho$ \\
\hline
sigma-map Union2.1 vs. sigma-map JLA & $+0.614$  \\ 
Hubble-map Union2.1 vs. Hubble-map JLA & $-0.188$   \\ 
q-map Union2.1 vs. q-map JLA & $+0.261$  \\  
\hline
\hline
\end{tabular}
\end{center}
\caption{ The correlation between pairs of maps. 
We have adopted the Pearson's coefficient $\rho$ to measure the correlation between 
diverse pairs of maps.} 
\end{table}


\subsection{The multipole analyses and Monte Carlo tests}

\begin{figure*}[!t]
\includegraphics[width = 6.0cm, height = 8.75cm, angle = -90]
{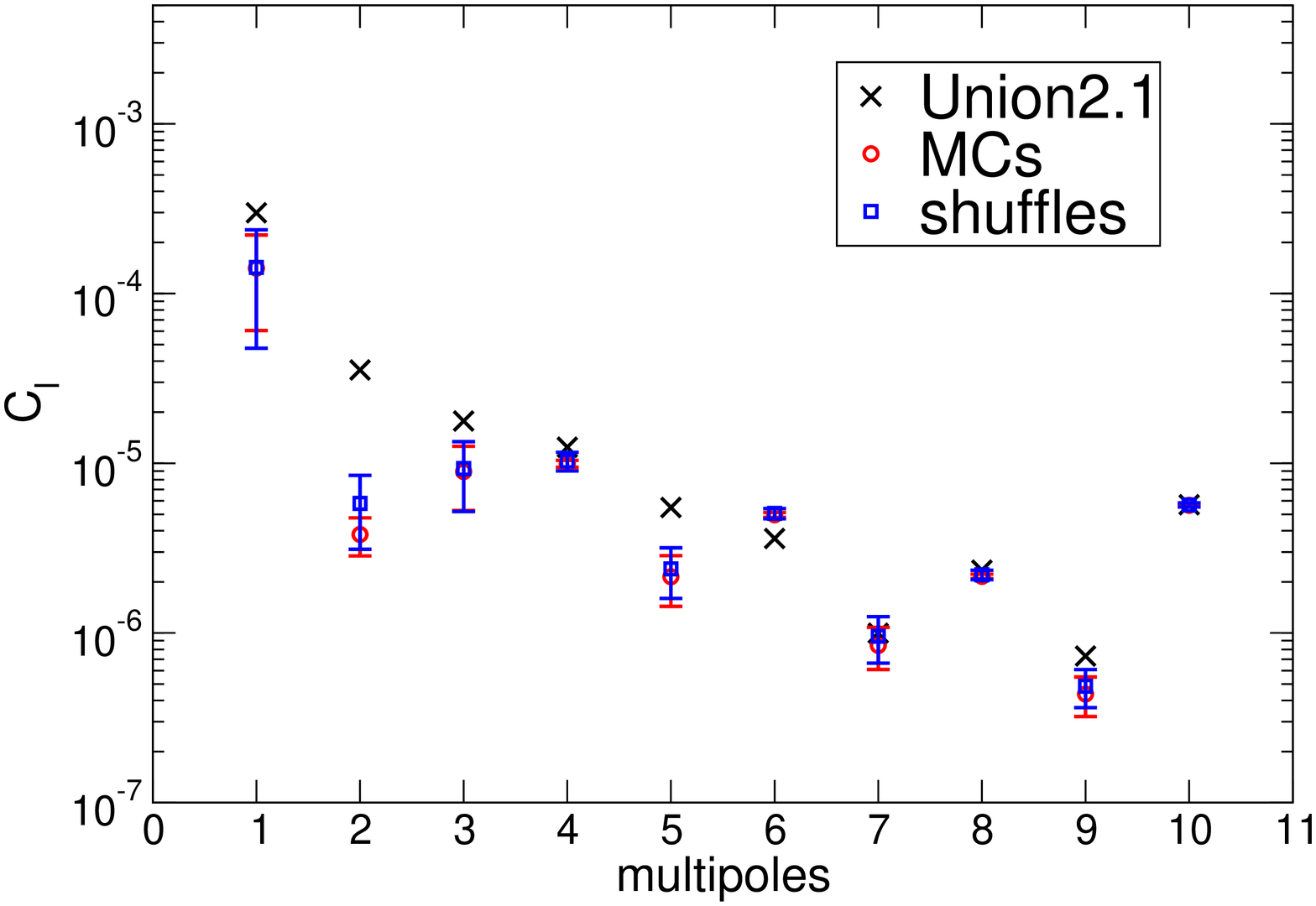}
\hspace{0.3cm}
\includegraphics[width = 6.0cm, height = 8.75cm, angle = -90]
{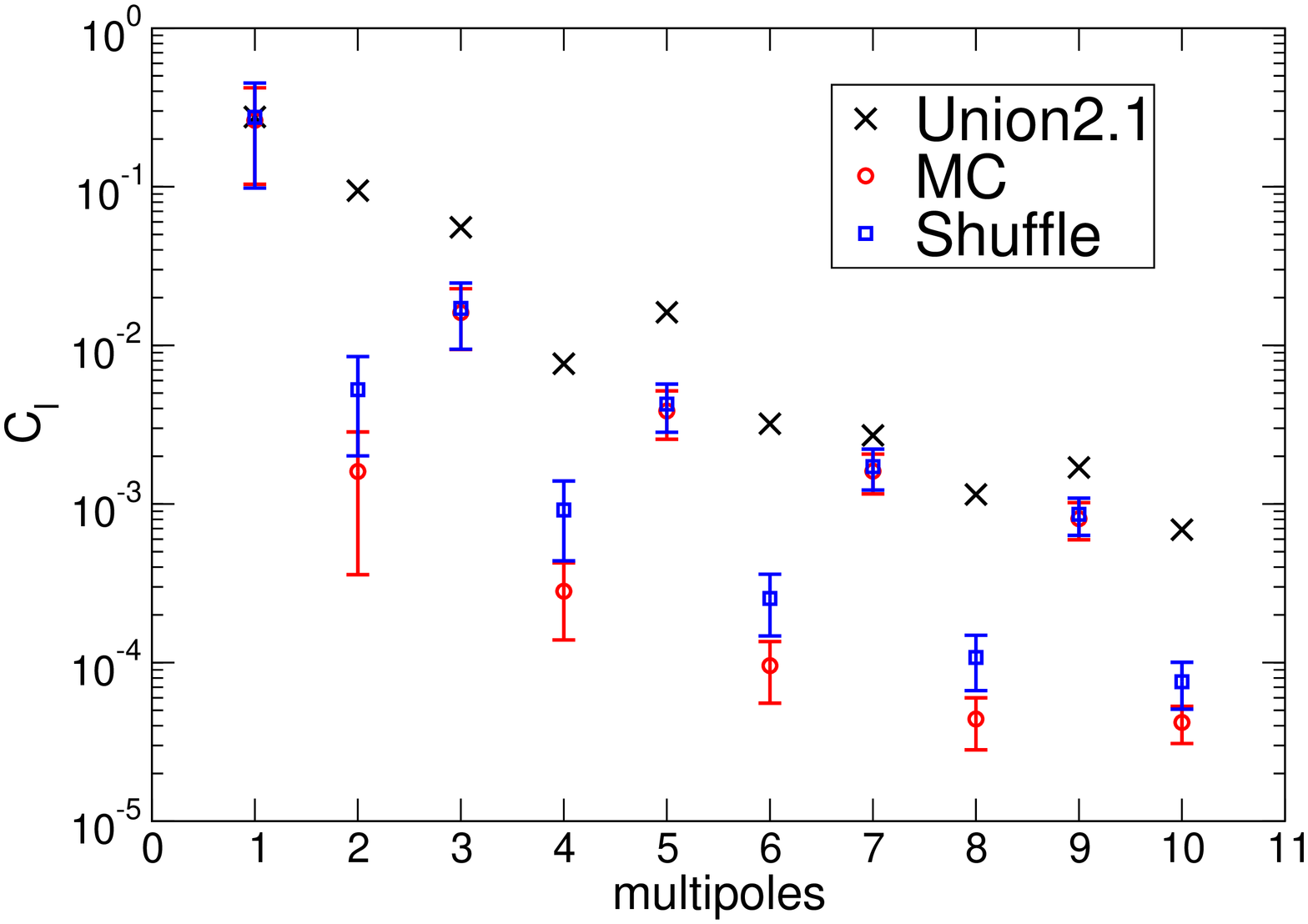}
\caption{{\it Left panel:} The angular power spectrum, $\{C_{\ell}\}$, up to $\ell = 10$, 
of the Hubble-map for the Union2.1 catalog. {\it Right panel:} the same for the q-map. 
The crosses represent the values for the original dataset, while the red (blue) circles (squares)
assign the average spectra from Hubble-maps obtained from 500 MC (shuffle) realisations.
Their respective error bars are estimated using the median absolute deviation of each coefficient of these spectra.
}
\label{fig10}
\end{figure*}

\begin{figure*}[!t]
\includegraphics[width = 6.0cm, height = 8.75cm, angle = -90]
{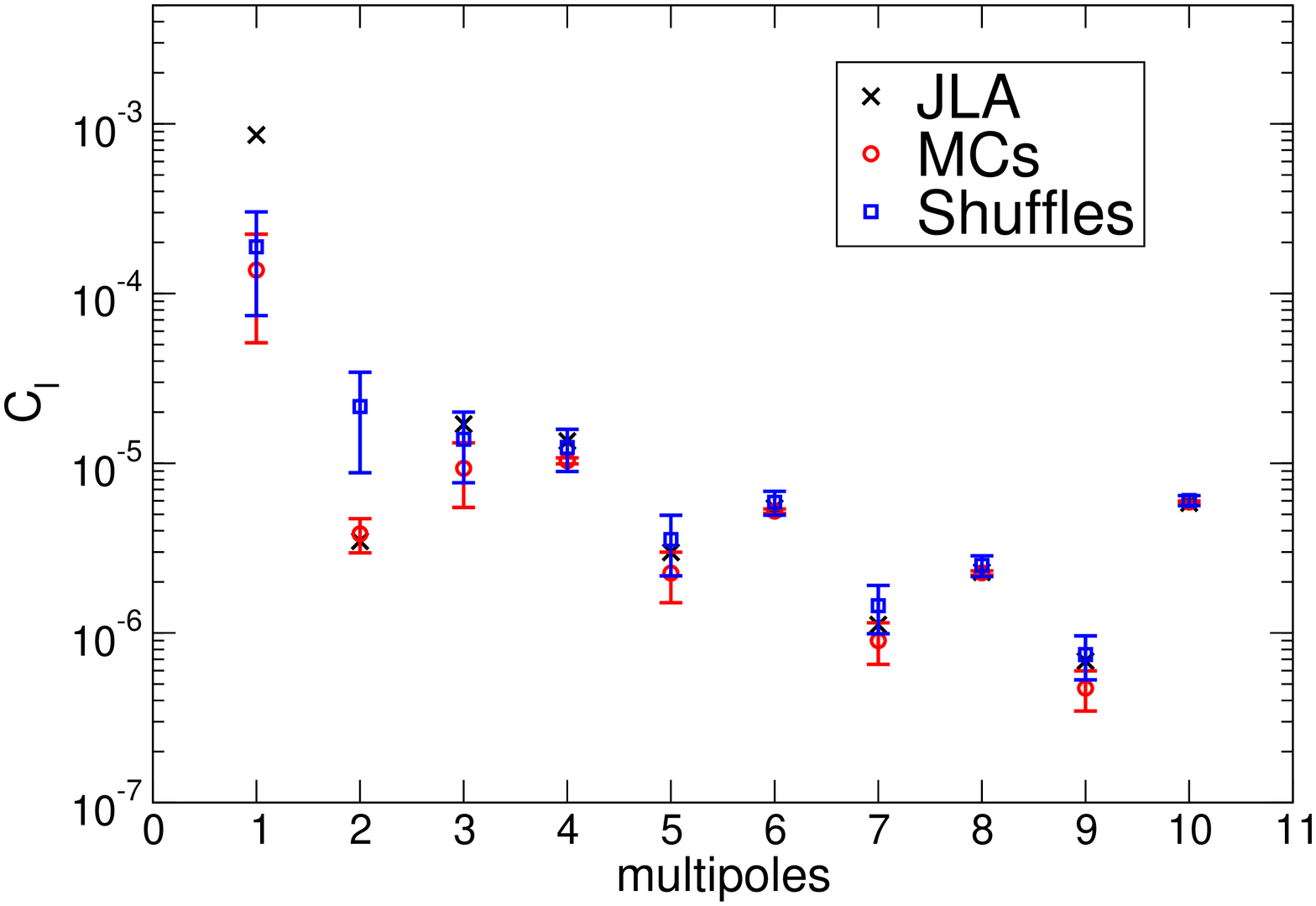}
\hspace{0.3cm}
\includegraphics[width = 6.0cm, height = 8.75cm, angle = -90]
{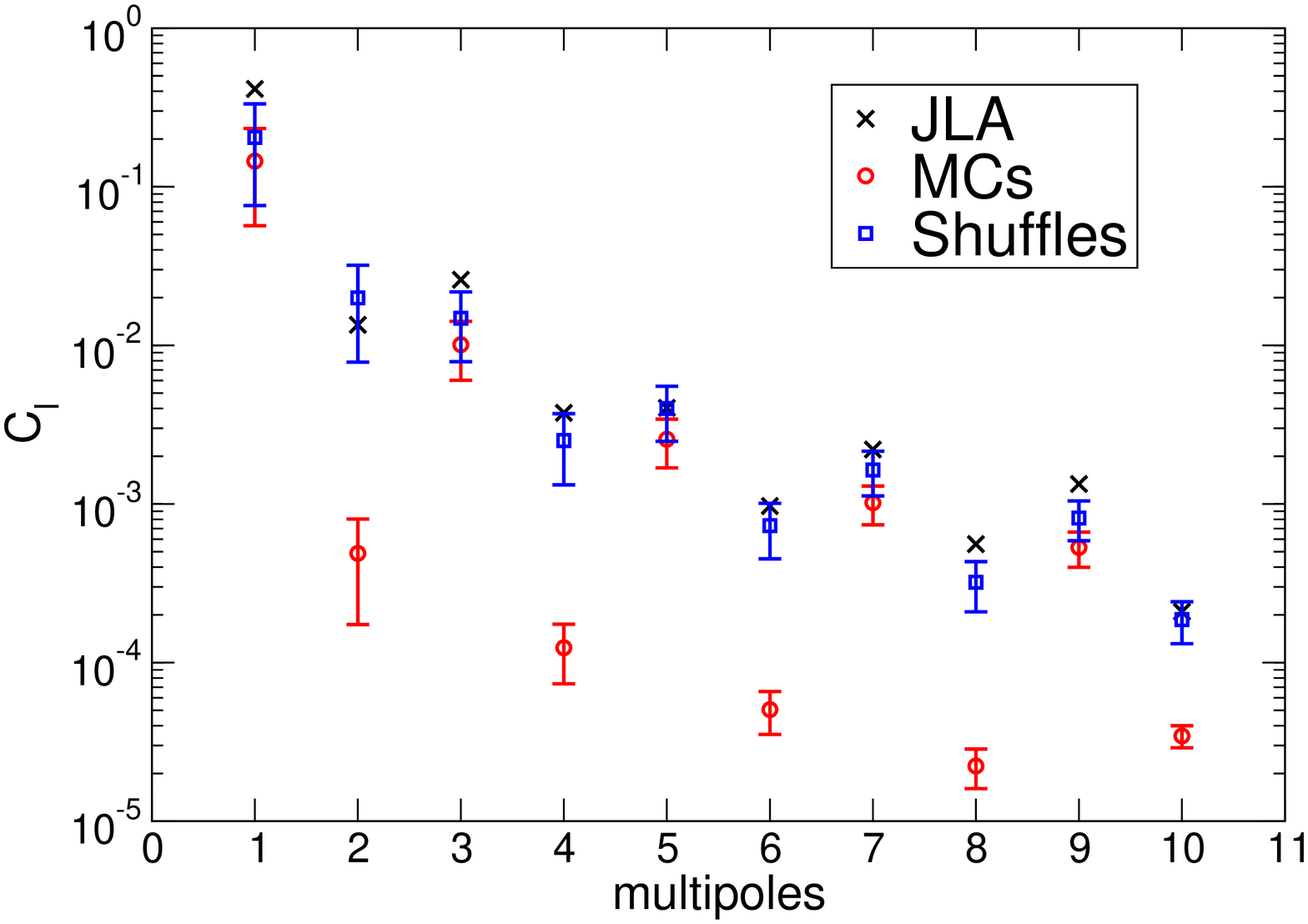}
\caption{{\it Left panel:} The angular power spectrum, $\{C_{\ell}\}$, up to $\ell = 10$, 
of the Hubble-map for the JLA catalog. {\it Right panel:} the same for the q-map. 
The crosses represent the values for the original dataset, while the red (blue) circles (squares)
assign the average spectra from Hubble-maps obtained from 500 MC (shuffle) realisations.
Their respective error bars are estimated using the median absolute deviation of each coefficient of these spectra.
}
\label{fig11}
\end{figure*}

The figures~\ref{fig10} and \ref{fig11} features the multipole expansion for the Hubble-maps 
(left panel) and $q$-maps (right panel) for the original Union2.1 and JLA datasets, respectively, 
compared to 500 realisations of MC and shuffle tests, where table 6 features the  
percentage of Shuffle and MC realisations whose $\delta H_0$ or $\delta q_0$ 
are larger than that of the the original datasets.


\begin{table}[!t]
\begin{center}
\label{tab:tab_shuff_mc} 
\begin{tabular}{cccc}
\hline
\hline
SNe catalogs & MCs (\%) & Shuffles (\%)\\
\hline
Union2.1 ($z \leq 0.2$) hubble-map & $3.0$ & $7.6$ \\
JLA ($z \leq 0.2$) hubble-map & $0.6$ & $18.2$ \\
\hline
Union2.1 ($z \leq 0.2$) q-map & $0.0$ & $0.0$\\
JLA ($z \leq 0.2$) q-map & $4.2$ & $42.8$ \\
\hline
\hline
\end{tabular}
\end{center}
\caption{ The percentage of Shuffle and MC realisations whose $\Delta h$ or $\Delta q_0$ 
are larger than that of the original datasets.} 
\end{table}


In the left of~\ref{fig10}, we can note that, for most $\ell$, the original data present larger 
$c_\ell$s than the simulated datasets except for the dipole coefficient, whose concordance
is mildly larger than 1$\sigma$ uncertainty. On the other hand, the asymmetry of $H_0$, 
$\delta H_0 = 0.046$,
of these simulations does not present significant disagreement with the result obtained from data, 
as 7.6\% of the shuffled realisations yield $\delta H_0\geq 0.046$ of the Union2.1, 
whereas this number decreases to 3.0\% for the MC test.  
By the same token, 18.2\% of the shuffle test exceed the asymmetry on JLA Hubble-map ($0.055$), 
however this percentage drops to 0.6\% when performing the MC realisations. This result also
shows the intrinsic non-uniformity of the angular SNe distribution of this sample, as the MC test
yields smaller asymmetric signals than the original map when we redistribute
the objects in the sky in uniform manner.

In the right panels of both figures, for instance, we stress the comparison 
between the $q$-maps and the simulated realisations of the Union2.1 and JLA data.
For the first compilation, the multipole expansion of the $q$-maps clearly indicates 
a significant asymmetry, as most $\ell$s (except for the dipole term) 
are not reproduced by the simulations we have produced, such that this asymmetry is reflected 
on the $\delta q_0$ computation, where none of the MC and shuffle realisations
exceed the value obtained for the original sample, that is $\delta q_0 = 2.56$.
Although this large asymmetry can not be reproduced by these statistical tests, 
we notice that this discrepancy is mostly due to the presence of an outlier fit for $q_0$, i.e., 
${q_0}_{\rm max} = +0.160$, and that no other hemisphere provide a positive fits
for the deceleration parameter except for this. 
Such large range of values allowed for $q_0$ can be attributed to a 
low constraining power on $q_0$ from this SNe dataset. Hence, it remains inconclusive 
whether the cosmic acceleration is actually anisotropic or not albeit the low statistical
significance of this result. 


Nevertheless, the JLA q-map presents a better agreement with these statistical tests, 
specially the shuffle test, as shown in the multipole expansion on the right plot of the figure~\ref{fig11},
in addition to the fact that 42.8\% of these shuffle realisations present $\delta q_0 \geq 1.62$. 
However, the MC test, show a smaller concordance with the data, as 
only 4.2\% of the simulations are able to reproduce such anisotropy on the q-map. 
As expected, the smaller asymmetry on idealistically isotropic JLA samples reinforces the impact of the 
incomplete sky coverage of these catalog.

It is also worth to discuss that the directions of the maximal and minimal $(H_0, q_0)$,
besides the sigma-map cases, are randomly oriented in both statistical tests, 
These positions are reproduced by less than 2\% of the 500 realisations
in all cases with a tolerance of $\sim \pm 11^{\circ}$ in both latitude and longitude coordinates.
Therefore, it is possible to conclude that such directions does not present
statistical significances, even though they coincide with other CP probes, 
and show suggestive concordance with the reports from the literature which 
have adopted similar methodologies. 

To summarise, we cannot conclude that the anisotropical signals found on the Union2.1 catalog 
analyses are influenced by the angular correlations of SNe distribution in the sky, together
with the reasonably low statistical significance of these results, however, the proximity
of the bulk-flow motion directions reported in the literature raises the possibility 
of being the cause for such signals. The JLA analyses, on the other hand, 
clearly show that its anisotropic coverage of the sky indeed 
impact the mapping of cosmological parameters through opposite hemispheres. 

Therefore, it is not possible to determine any violation of the CP, in this redshift range,
with the limitation of the current datasets. 
Nevertheless, next-generation surveys such as LSST (Abell {\it et al.}, 2009)
and Euclid (Amendola {\it et al.}, 2012) may solve this issue with the greater
precision, and much larger datasets, that they shall provide.

\section{Conclusions}

In this work we have discussed the issue of a possible anisotropic expansion of the Universe through a directional analyses, i.e., investigating the Hubble diagrams of SNe data located in opposite hemispheres. For completeness, our analyses also explore the possibility that the results are biased by the anisotropic 
distribution of the SNe data, because they are manifestly under-sampled or over-sampled in several 
sky patches. For this, we have firstly used a geometrical method to find possible anisotropic signatures 
due to the angular distribution of both SNe datasets, namely the Union2.1 and the JLA catalogs. 
Then we use this information to investigate a possible correlation with the outcomes of the directional 
studies of $H_0$ and $q_0$. 

In fact, according to our analyses, a correlation exists between the anisotropic distribution of the data 
ensambles and the anisotropic signatures found in our directional studies of $H_0$ and $q_0$. 
Specifically, the JLA data shows a large correlation that is statistically significant to suggest that the 
dipolar direction of the Hubble-map can be attributed to the anisotropically distributed SNe sample. 
In the case of the Union2.1 catalog, this correlation is small. 
However, we have verified that the direction of the dipole asymmetry found in the $H_0$ analysis 
with this dataset correlates well with the well-known direction of the bulk-flow motion, in fact 
we obtained $(l,b) = (326^{\circ}, 12^{\circ})$ while the bulk-flow motion of our local group points 
towards $(l,b) = (319^{\circ}, 7^{\circ})$ (Turnbull {\it et al.}, 2012). 

All these results indicate that the dipolar features of the directional Hubble diagrams (which is the 
dominant anisotropic signal) can be explained either by the anisotropic SNe dataset (in the case of the 
JLA sample) or by a systematic effect, that is, the direction of the bulk-flow motion of our local group in 
the case of the Union2.1 catalog. 
Thus, for the low-redshift regime of the SNe analysed here ($z \le 0.2$), we have shown that the observed anisotropic expansion cannot guarantee a violation of the CP. 
Despite of this, one cannot discard the hypothesis that our Universe is indeed undergoing an anisotropic 
expansion, which is not observed in the analyses because either it could be of small magnitude or it may be hidden 
by the uncertainties introduced by the highly non-uniform distribution of current SNe data in the celestial sphere. 

\vspace{8.0cm}

\acknowledgements
We are grateful to Roy Maartens, Ribamar Reis and Ivan Soares Ferreira for helpful discussions. 
We acknowledge CNPq, FAPERJ and CAPES for financial support.


\label{lastpage}

\end{document}